\documentclass[aps, prd, reprint, superscriptaddress, amsmath, amssymb, nofootinbib, floatfix]{revtex4-1}

\usepackage{graphicx}
\usepackage[table,xcdraw]{xcolor}
\usepackage[caption=false]{subfig}
\usepackage{multirow}
\usepackage{booktabs}
\usepackage[normalem]{ulem} 

\usepackage{bm}
\usepackage{physics}
\usepackage{slashed}
\usepackage{dcolumn}

\newcommand{\nubarLamNucMin}{3}
\newcommand{\nubarLamNucMax}{300}
\newcommand{\nubarLamDMin}{0.5}
\newcommand{\nubarLamDMax}{16}
\newcommand{\nubarLamHiYLmin}{0.46}
\newcommand{\nubarLamHiYLmax}{4}
\newcommand{\nubarLamLoYLmin}{1.8}
\newcommand{\nubarLamLoYLmax}{16}
\newcommand{\nubarRatioMin}{83}
\newcommand{\nubarRatioMaxExp}{4}
\newcommand{\nubarRatioMaxMant}{4.5}
\newcommand{\smallYDratio}{1600}
\newcommand{\xoneLamMin}{0.7}
\newcommand{\xoneLamMax}{9}
\newcommand{\nuRatioYLthreeMax}{3.9}
\newcommand{\nuRatioYLhiMax}{6.5}
\newcommand{\nuRatioYLoneMin}{4}
\newcommand{\nuRatioYLoneMax}{200}
\newcommand{\DtwoRedHi}{32}
\newcommand{\DtwoRedLow}{3.0}
\newcommand{\DtwoRedYLtwo}{1.7}
\newcommand{\nubarTwoBodyMax}{2.4}
\newcommand{\GamMedMin}{4.5}
\newcommand{\GamMedMax}{40}
\newcommand{\GamRefMin}{7}
\newcommand{\GamRefMax}{23}
\newcommand{\GamRatioMin}{0.57}
\newcommand{\GamRatioMax}{1.7}%
\newcommand{\gcritTr}{0.60--0.75}
\newcommand{\gcritTrSc}{0.85--1.03}
\newcommand{\gcritNf}{0.68--0.78}
\newcommand{\gcritNfSc}{0.88--1.05}
\newcommand{\gcritNuc}{0.65--1.08}
\newcommand{\gcritNucSc}{0.90--1.28}
\newcommand{\mupiNucLoYL}{164--207}
\newcommand{\mupiDLoYL}{67--141}
\newcommand{\chenRatios}{1.09, 1.72, 2.4 and 4.2}
\newcommand{\chenLam}{0.20, 0.14, 0.12 and 0.11}
\newcommand{\chenFactor}{2.3--2.4}
\newcommand{\chenTheirRatios}{1.07, 1.51, 2.3 and 5.5}

\usepackage{url}
\usepackage[colorlinks=true, allcolors=blue]{hyperref}

\definecolor{violet}{rgb}{102, 0, 153}

\begin{document}
\title{$\Delta(1232)$ isobars and neutrino opacities in hot dense matter}
\author{Subha Mitra}
\affiliation{Department of Physics \& Astronomy, National Institute of Technology, Rourkela 769008, India}

\author{Bharat Kumar}
\email{kumarbh@nitrkl.ac.in}
\affiliation{Department of Physics \& Astronomy, National Institute of Technology, Rourkela 769008, India}

\date{\today}

\begin{abstract}
We calculated the diffusion mean free paths of electron neutrinos and antineutrinos in hot, neutrino-trapped matter containing $\Delta(1232)$ isobars, as found in proto-neutron stars and neutron star merger remnants. We included charged-current absorption with the $N\Delta$ weak vertex of weak pion production, neutral-current scattering on all baryons and three-body antineutrino absorption on electron--baryon pairs, with relativistic mean-field kinematics. The reaction $\bar\nu_e+e^-+n\to\Delta^-$ dominates the antineutrino opacity whenever the neutron and electron chemical potentials exceed the in-medium $\Delta^-$ rest energy, even where the $\Delta$ population is negligible. At two to four times saturation density, where the reaction is open, it shortens the antineutrino mean free path from \nubarLamNucMin--\nubarLamNucMax~m to \nubarLamDMin--\nubarLamDMax~cm. Because antineutrinos are scarce at high lepton fraction, this changes lepton-number diffusion only at low lepton fraction, by up to \DtwoRedHi\% at these densities and by a factor of \DtwoRedLow\ at 1.5 times saturation density. For electron neutrinos, absorption on the $\Delta^-$ and, at low lepton fraction, the opening of nucleonic absorption shorten the mean free path by factors of up to \nuRatioYLthreeMax\ at lepton fraction 0.3 and up to about \nuRatioYLoneMax\ at lepton fraction 0.1. We also showed that, for a given axial coupling, the spin of the $\Delta$ reduces its Gamow--Teller strength per particle from 3 to $5/3$.
\end{abstract}

\maketitle

\section{Introduction}
\label{sec:intro}
Neutrinos carry away the lepton number and the thermal energy of a proto-neutron star during the first tens of seconds after core collapse~\cite{BurrowsLattimer1986,Pons:1998mm}. In neutron star merger remnants they are the main source of cooling~\cite{Foucart2023} and drive changes in the composition of the remnant and of the ejecta~\cite{Foucart2023,Sedrakian:2026lfd}. In both systems the inner regions reach densities of a few times the nuclear saturation density $n_0$, and temperatures of tens of MeV are reached in parts of them~\cite{Prakash:1996xs,Perego2019,Alford:2026kwd}. There the neutrinos are trapped: their mean free paths are short compared to the size of the hot region, they form a gas close to equilibrium with the matter, and their transport is diffusive. In mergers trapping becomes important above temperatures of about 5--10~MeV~\cite{AlfordHarris2018}, and a mean-free-path estimate places their thermalization above about 20~MeV, while Monte Carlo transport finds thermal distributions in all observables only from about 30~MeV~\cite{Alford:2026kwd}. The rate of diffusion is set by the opacity of the dense matter. Proto-neutron star evolution calculations with opacities computed consistently with the equation of state showed that the deleptonization and cooling timescales depend on these opacities~\cite{Pons:1998mm}. In merger remnants electron antineutrinos are the most abundant trapped species in the high-density regions, with number fractions of order $10^{-3}$~\cite{Espino2024}. The antineutrino opacity therefore matters for the transport of lepton number in mergers, and it is the quantity most sensitive to the new reaction we discuss below.

The opacities of nucleonic matter have been calculated with increasing care. Mean-field shifts of the single-particle energies change the kinematics of charged-current absorption~\cite{RPL1998,MartinezPinedo2012,RobertsReddyShen2012}, the full relativistic response contains terms omitted in earlier derivations~\cite{RobertsReddy2017}, and correlations reduce the cross sections~\cite{BurrowsSawyer1998,RPLP1999} and redistribute the strength of the charged-current response~\cite{Shin:2023sei}. The mean-field shifts change the spectra and luminosities of the emitted neutrinos~\cite{MartinezPinedo2012}, and so does the full reaction kinematics~\cite{Fischer2020}. Residual interactions in the spin and spin-isospin channels have a strong effect on the opacities and remain poorly constrained, as shown up to about saturation density~\cite{Lin:2022lug}, and magnetic fields of up to $10^{17}$~G enhance the charged-current opacities of low-energy neutrinos at subnuclear densities~\cite{Kumamoto:2026aek}. For antineutrinos an additional channel is the inverse neutron decay $\bar\nu_e+e^-+p\to n$, an absorption on an electron--proton pair. It was included in proto-neutron star evolution in Ref.~\cite{Fischer2020}, it is the dominant charged-current process for low-energy antineutrinos under neutrinosphere conditions~\cite{Fischer2020,Oertel2020}, and it is now part of opacity sets used in simulations~\cite{Oertel2020,Pascal2022}. Heavier hadrons open further channels. The $\Sigma^-$ and $\Xi^-$ hyperons, which have large weak vector charges, contribute significantly to neutrino scattering, and in lepton-rich matter hyperons also raise the neutrino energies and thereby the cross sections~\cite{Reddy:1996tw}. At a representative proto-neutron star condition the $\Lambda$ hyperons change the electron-flavor opacities little, because their abundance is small, their charged-current transitions are Cabibbo suppressed and the nucleonic electron-flavor absorption has no kinematic threshold; they matter mainly for low-energy muon neutrinos and antineutrinos, for which they open absorption channels including the three-body reaction $p+\mu^-+\bar\nu_\mu\to\Lambda$~\cite{Zatini2026}. Thermal pions open further charged-current reactions of muon neutrinos~\cite{ForeReddy2020}.

The $\Delta(1232)$ isobars are the lightest baryon resonances~\cite{PDG2024}, and their weak transitions to nucleons are Cabibbo allowed. In vacuum they decay into a nucleon and a pion; in dense matter this decay is expected to be Pauli blocked~\cite{RadutaOertelSedrakian2020,Raduta2021}, and in neutrino-transparent matter up to 10~MeV the populated $\Delta$s were found to be long-lived quasiparticles in the mean-field description~\cite{Kumar2026P1}. The $\Delta^-$ appears first, because it replaces energetic electrons as the carrier of negative charge~\cite{SedrakianLiWeber2023}. In relativistic mean-field models with near-universal couplings it appears in cold matter at about two to three times $n_0$, depending on the slope of the symmetry energy and on the $\rho$--$\Delta$ coupling~\cite{Drago2014,Cai2015}, and close to saturation density if the $\Delta$ potential in matter is sufficiently attractive~\cite{LiSedrakianWeber2018}. For a range of $\Delta$--meson couplings the maximum mass remains above two solar masses~\cite{Kolomeitsev2017,LiSedrakianWeber2018,Raduta2021}, although in some models a $\Delta$ potential close to the empirical one softens the equation of state too much~\cite{Drago2014}. The $\Delta$ potential itself is constrained only loosely; pion production in heavy-ion collisions favors an isoscalar potential moderately more attractive than the nucleon one~\cite{CozmaTsang2021}. At finite temperature and fixed lepton fraction the $\Delta$s appear at lower densities~\cite{RadutaOertelSedrakian2020}; in the models of Refs.~\cite{Malfatti2019,RadutaOertelSedrakian2020} all four charge states are populated at a few times $n_0$, with the $\Delta^-$ the most abundant~\cite{RadutaOertelSedrakian2020}. In neutrino-trapped matter at several tens of MeV they make up of order ten percent of the baryons, which led Sedrakian and Harutyunyan to suggest that they could be a new source of neutrino opacity~\cite{SedrakianHarutyunyan2022}.

Neutrino opacities of $\Delta$-admixed matter were first computed by Chen \textit{et al.} in the GM3 relativistic mean-field model, at $T=30$~MeV and, for trapped neutrinos, a lepton fraction $Y_L=0.4$. For neutral-current scattering they found that the large vector coupling of the $\Delta^-$ and the high spin of the $\Delta$ make the $\Delta^-$ dominate the scattering rate, and that in trapped matter the $\Delta$s shorten the neutrino mean free path drastically~\cite{Chen2007}; Ref.~\cite{ChenGuo2008} also reported a rapid decrease of the neutrino mean free paths due to the $\Delta$s. Adding charged-current reactions of electron neutrinos, with the axial coupling of $\nu_e+\Delta^0\to e^-+\Delta^+$ set to one, the other $\Delta$ axial couplings scaled from it, and an $N\Delta$ vertex without $q^2$ dependence, they found the charged-current and neutral-current contributions comparable, and the transitions between nucleons and $\Delta$s to contribute at most 20\% of the charged-current opacity in neutrino-free matter~\cite{Chen2009}. They did not compute antineutrino opacities. The $\Delta$s also modify weak processes in neutrino-transparent matter. In a companion paper we showed that the $\Delta^-$ takes over the negative charge from the leptons, raises the proton fraction and opens the nucleonic direct Urca process in models where it is otherwise closed~\cite{Kumar2026P1}, and the emissivities of $\Delta$ direct Urca processes in cold neutron stars were reported to be comparable to those of nucleonic direct Urca~\cite{Yogeesh2026}.

Two features of $\Delta$-admixed matter have not been examined. First, the three-body absorption that complements the inverse neutron decay in nucleonic matter has a $\Delta$ analogue, $\bar\nu_e+e^-+n\to\Delta^-$. In $\beta$ equilibrium with trapped neutrinos $\mu_{\Delta^-}=\mu_n+\mu_e-\mu_{\nu_e}$, so this reaction is open whenever $\mu_n+\mu_e$ exceeds the in-medium $\Delta^-$ rest energy, even where $\mu_{\Delta^-}$ lies below it and the $\Delta^-$ population is negligible: its threshold is set by the $\Delta$ potential, not by the $\Delta$ abundance. Second, the $\Delta^-$ produced in this way carries momenta far above its Fermi surface, where its decay into a nucleon and a pion is not Pauli blocked. Its width is then controlled by the pion propagator in the medium, which depends on the short-range spin-isospin interactions described by Landau--Migdal parameters~\cite{Migdal1990,HelgessonRandrup1995}. With the values of these parameters inferred from Gamow--Teller quenching, $p$-wave pion condensation has been predicted at densities as low as 1.6--1.7 times saturation density~\cite{SuzukiSakaiTatsumi1999}, so the stability of the uncondensed matter must itself be checked.

In this paper we calculated the opacities of hot, neutrino-trapped matter containing the $\Delta$ quartet for electron neutrinos and antineutrinos, under conditions of proto-neutron stars and merger remnants. Our calculation has six ingredients. First, we included all charged-current channels, $N\to N$, $\Delta\to N$, $N\to\Delta$ and $\Delta\to\Delta'$, with relativistic Rarita--Schwinger matrix elements and the $N\Delta$ weak transition vertex of Ref.~\cite{HNV2007} without its pion-pole term (bounded separately in Sec.~\ref{sec:opac}), crossed to absorption, and neutral-current scattering on every member of the quartet, all with the full relativistic phase space and mean-field kinematics. Second, we included three-body antineutrino absorption on an electron and a baryon for every charged-current pair, which contains the inverse neutron decay of Refs.~\cite{Fischer2020,Oertel2020} and, with $\Delta$s, $\bar\nu_e+e^-+n\to\Delta^-$. Third, we derived the static limit of neutrino scattering on a spin-$\tfrac32$ quasiparticle, which shows how the spin of the $\Delta$ enters the cross section. Fourth, we computed the diffusion mean free path in matter in $\beta$ equilibrium with trapped neutrinos at fixed lepton fraction, separated the direct contribution of the $\Delta$ reactions from the indirect one through the change of composition, and evaluated the antineutrino share of lepton-number diffusion. Fifth, we computed the width of the produced $\Delta^-$ with an in-medium pion self-energy and determined the range of Landau--Migdal parameters for which the matter is stable against pion condensation. Sixth, we tested the results against collisional broadening, long-wavelength correlations, a second equation of state, the uncertain $\Delta$ weak couplings and the earlier calculation of Ref.~\cite{Chen2007}. We are not aware of an earlier treatment of $\bar\nu_e+e^-+n\to\Delta^-$ in dense matter; its closest analogue is the hyperonic three-body absorption $p+\ell^-+\bar\nu_\ell\to\Lambda$ ($\ell=e,\mu$) of Ref.~\cite{Zatini2026}.

We found that $\bar\nu_e+e^-+n\to\Delta^-$ raises the antineutrino opacity at 2--$4n_0$ by two to four orders of magnitude, also in matter with a negligible $\Delta$ population, and that it lowers the lepton-number diffusion coefficient at low lepton fraction by up to \DtwoRedHi\%, and by a factor of \DtwoRedLow\ at $1.5n_0$. For electron neutrinos, except at low lepton fraction, the enhancement of the opacity is set by the total $\Delta$ fraction. The in-medium width of the produced $\Delta^-$ is \GamMedMin--\GamMedMax~MeV, and the matter, with or without $\Delta$s, is stable against pion condensation only if the $\Delta\Delta$ Landau--Migdal parameter exceeds a critical value between about 0.6 and 1.3, depending on the state and on the normalization of the $p$-wave attraction.

Section~\ref{sec:matter} describes the matter models and the equilibrium with trapped neutrinos. Section~\ref{sec:opac} derives the opacities: the reactions, the phase space of two- and three-body absorption, the matrix elements and their crossing, the static limit for spin-$\tfrac32$ targets, the diffusion mean free path and the many-body corrections. Section~\ref{sec:valid} describes the numerical checks, Sec.~\ref{sec:results} presents the composition, the opacities, the diffusion mean free paths, lepton-number transport, the in-medium width of the $\Delta^-$ and the stability against pion condensation, and Sec.~\ref{sec:concl} discusses the implications. The phase-space reduction, the spin-summed functions, the width treatments, the benchmark against Ref.~\cite{Chen2007} and the numerical convergence are given in Appendices~\ref{app:ps}--\ref{app:conv}.

\section{Matter with $\Delta$ isobars and trapped neutrinos}
\label{sec:matter}
We use natural units and the metric signature $(+,-,-,-)$, and convert to CGS units only for mean free paths. Baryon densities are given in units of $n_0=0.16$~fm$^{-3}$, which we use as a fixed unit for both models; their saturation densities are 0.152~fm$^{-3}$ (DDME2) and 0.157~fm$^{-3}$ (QMC-RMF3)~\cite{DDME2,ABHT2022}.

\subsection{Models}
We use the two relativistic mean-field models of Ref.~\cite{Kumar2026P1}. DDME2~\cite{DDME2} has density-dependent meson--nucleon couplings fitted to nuclear properties. QMC-RMF3~\cite{ABHT2022} has constant couplings with nonlinear meson self-interactions, fitted to chiral effective field theory calculations of neutron matter and to the properties of symmetric matter at saturation. Both are extended by the $\Delta$ quartet, coupled to the mesons through the ratios $x_{\sigma\Delta}$, $x_{\omega\Delta}$ and $x_{\rho\Delta}$ of the $\Delta$ to the nucleon couplings. Neither model allows the nucleonic direct Urca process in cold, neutrino-free $\beta$-equilibrated matter at the densities reached in neutron stars~\cite{RadutaOertelSedrakian2020,ABHT2022}, so the opening of nucleonic absorption that we find below is caused by the $\Delta$s. The Lagrangian, the field equations and the finite-temperature thermodynamics are those of Ref.~\cite{Kumar2026P1}. DDME2 is our reference model; QMC-RMF3 serves as a test of the dependence on the equation of state.

We take $x_{\omega\Delta}=1.1$, $x_{\rho\Delta}=1.0$ and $x_{\sigma\Delta}=1.1$ or 1.2 (Table~\ref{tab:inputs}). The first set was used for DDME2 in Ref.~\cite{RadutaOertelSedrakian2020}, where it gives a $\Delta$ potential of about $-83$~MeV in symmetric matter at saturation ($-85$~MeV in our solver, with a nucleon potential $U_N=-77$~MeV). This lies within the range $U_N-30~\text{MeV}\lesssim U_\Delta\lesssim U_N$ inferred from electron and pion scattering on nuclei and from photoabsorption~\cite{Drago2014,SedrakianLiWeber2023}, with $U_N$ the nucleon potential; other, partly conflicting estimates are discussed in Ref.~\cite{Kolomeitsev2017}, and pion production in heavy-ion collisions favors an isoscalar potential moderately more attractive than the nucleon one~\cite{CozmaTsang2021}. Raising $x_{\sigma\Delta}$ to 1.2 makes the $\Delta$ potential about 40~MeV more attractive, beyond that range; this value lies within the coupling ranges explored in Refs.~\cite{LiSedrakianWeber2018,SedrakianLiWeber2023}, and we use it as a case with a large $\Delta$ fraction.

\subsection{Equilibrium with trapped neutrinos}
Strong equilibrium among the baryons fixes all baryon chemical potentials in terms of the baryon and charge chemical potentials,
\begin{equation}
\mu_b=\mu_B+q_b\,\mu_Q ,
\label{eq:strong}
\end{equation}
with $q_b$ the electric charge of baryon $b$. With trapped electron neutrinos, weak equilibrium and the conservation of lepton number read
\begin{equation}
\mu_n-\mu_p=\mu_e-\mu_{\nu_e},\qquad Y_e+Y_{\nu_e}=Y_L ,
\label{eq:weak}
\end{equation}
where $Y_e$ and $Y_{\nu_e}$ are the net electron and net electron-neutrino numbers per baryon and $Y_L$ is the lepton fraction. Charge neutrality requires, with $n_e$ the net (electron minus positron) density,
\begin{equation}
\sum_b q_b\,n_b=n_e .
\label{eq:neutral}
\end{equation}
The net density of a massless neutrino species with one helicity state is
\begin{equation}
n_{\nu_e}-n_{\bar\nu_e}=\frac{1}{6\pi^2}\Big(\mu_{\nu_e}^3+\pi^2T^2\mu_{\nu_e}\Big) ,
\label{eq:nnu}
\end{equation}
which holds at any temperature. These are the equilibrium conditions of lepton-rich proto-neutron star matter~\cite{Prakash:1996xs,RPL1998}. We neglect muons and set the muon lepton number to zero, as in Refs.~\cite{Malfatti2019,RadutaOertelSedrakian2020}, and we do not consider $\nu_\mu$ and $\nu_\tau$. For given $(n_B,T,Y_L)$ we solve the mean-field equations at fixed charge fraction and adjust $Y_e$ until Eq.~\eqref{eq:weak} holds. The equilibrium conditions are satisfied to better than $10^{-7}$ relative accuracy.

\section{Neutrino opacities}
\label{sec:opac}
\subsection{Reactions}
The charged-current reactions are
\begin{equation}
\nu_e+L\to H+e^-,\qquad \bar\nu_e+H\to L+e^+ ,
\label{eq:cc}
\end{equation}
where $(L,H)$ is a pair of baryons whose charges differ by one: $(n,p)$, $(\Delta^-,n)$, $(\Delta^0,p)$, $(n,\Delta^+)$, $(p,\Delta^{++})$, $(\Delta^-,\Delta^0)$, $(\Delta^0,\Delta^+)$ and $(\Delta^+,\Delta^{++})$. The neutral-current reactions are elastic scattering on every baryon,
\begin{equation}
\nu+B\to\nu+B,\qquad \bar\nu+B\to\bar\nu+B .
\label{eq:nc}
\end{equation}
The channels, their isospin factors and their couplings are listed in Table~\ref{tab:channels}. Neutral-current $N\leftrightarrow\Delta$ transitions, $\nu+n\leftrightarrow\nu+\Delta^0$ and $\nu+p\leftrightarrow\nu+\Delta^+$ and their antineutrino analogues, proceed through the isovector current only. We evaluated them with the vertex of Eq.~\eqref{eq:GND}, the vector form factors multiplied by $1-2\sin^2\theta_W$, the isospin factor of $n\to\Delta^+$ and the coupling $G_F$, with each crossing checked against explicit spinor sums. For $x_{\sigma\Delta}=1.2$ at 2--$3n_0$, $T=10$ and 20~MeV and $Y_L=0.1$ and 0.3 they shorten $\lambda_R$ by at most 4.5\% for neutrinos and 0.3\% for antineutrinos. We omit them in the rest of the paper. We also neglect neutrino--electron scattering, which matters mainly for energy exchange, whereas lepton-number transport is dominated by the reactions with baryons~\cite{RPL1998}. Finally, we include the three-body absorption
\begin{equation}
\bar\nu_e+e^-+H\to L
\label{eq:id}
\end{equation}
for all pairs $(L,H)$. For $(L,H)=(n,p)$ this is inverse neutron decay, which in proto-neutron star evolution dominates the opacity of low-energy antineutrinos~\cite{Fischer2020}; the analogous coalescence $\bar\nu_e+e^-+p\to\Lambda$ is Cabibbo suppressed~\cite{Zatini2026}. The $\Delta$ reactions include $\bar\nu_e+e^-+n\to\Delta^-$ and $\bar\nu_e+e^-+p\to\Delta^0$, the inverses of the weak decays of the $\Delta^-$ and $\Delta^0$. Neutrinos have no such channel, since it would need a positron in the initial state. The inverse reactions, the weak decays $L\to H+e^-+\bar\nu_e$, emit antineutrinos; they enter the diffusion coefficient through detailed balance (Sec.~\ref{sec:diff}).

\begin{table}[tbp]
\caption{Weak channels. Charged current: isospin factor $|\mathcal I|^2$, normalized to one for $n\to p$~\cite{Kumar2026P1}, and the baryonic vertex. Neutral current: weak vector charge $c_V=T_3-2\sin^2\theta_W Q$ and axial charge $c_A=g\,T_3$, with $g=g_A$ for nucleons and $g=g_{A\Delta}$ for $\Delta$s, evaluated with the inputs of Table~\ref{tab:inputs}.}
\label{tab:channels}
\begin{ruledtabular}
\begin{tabular}{lcc}
Charged current $(L,H)$ & $|\mathcal I|^2$ & vertex \\
\hline
$(n,p)$ & 1 & $\gamma^\mu(1-g_A\gamma_5)$ \\
$(\Delta^-,n)$, $(p,\Delta^{++})$ & 3 & $\Gamma^{\sigma\mu}$, Eq.~\eqref{eq:GND} \\
$(\Delta^0,p)$, $(n,\Delta^+)$ & 1 & $\Gamma^{\sigma\mu}$, Eq.~\eqref{eq:GND} \\
$(\Delta^-,\Delta^0)$, $(\Delta^+,\Delta^{++})$ & 3 & $-g^{\alpha\beta}\gamma^\mu(1-g_{A\Delta}\gamma_5)$ \\
$(\Delta^0,\Delta^+)$ & 4 & $-g^{\alpha\beta}\gamma^\mu(1-g_{A\Delta}\gamma_5)$ \\
\hline
Neutral current & $c_V$ & $c_A$ \\
\hline
$n$ & $-0.5$ & $-0.63$ \\
$p$ & $0.038$ & $0.63$ \\
$\Delta^-$ & $-1.038$ & $-1.134$ \\
$\Delta^0$ & $-0.5$ & $-0.378$ \\
$\Delta^+$ & $0.038$ & $0.378$ \\
$\Delta^{++}$ & $0.575$ & $1.134$ \\
\end{tabular}
\end{ruledtabular}
\end{table}

\begin{table}[tbp]
\caption{Inputs: weak couplings, $\Delta$ meson couplings and the width and broadening parameters. The sensitivity ranges are explored in Sec.~\ref{sec:sens}.}
\label{tab:inputs}
\begin{ruledtabular}
\begin{tabular}{lll}
Quantity & Value & Range / source \\
\hline
$G_F$ (MeV$^{-2}$) & $1.166\times10^{-11}$ & \cite{PDG2024} \\
$\cos\theta_C$ & $0.974$ & $|V_{ud}|$~\cite{PDG2024} \\
$\sin^2\theta_W$ & $0.2312$ & $\overline{\rm MS}$ at $M_Z$~\cite{PDG2024} \\
$g_A$ & $1.26$ & exp.\ $1.2754$~\cite{PDG2024} \\
$C_5^A(0)$ & $1.2$ & $0.867$--$1.2$~\cite{HNV2007} \\
$C_3^V,C_4^V,C_5^V$ at $q^2=0$ & $2.13,\,-1.51,\,0.48$ & \cite{HNV2007} \\
$g_{A\Delta}$ & $0.6\,g_A$ & $0$--$g_A$ \\
$x_{\sigma\Delta}$ & $1.1$, $1.2$ & \cite{RadutaOertelSedrakian2020,Kumar2026P1} \\
$x_{\omega\Delta}$, $x_{\rho\Delta}$ & $1.1$, $1.0$ & \cite{RadutaOertelSedrakian2020,Kumar2026P1} \\
Collisional width & $T^2/T_W$ & $T_W=5$~MeV~\cite{AHZ2024nwa} \\
NC broadening FWHM & $T^2/T_W$ & this work \\
Final $\Delta$, 3-body & $T^2/T_W$; 117~MeV & vacuum~\cite{PDG2024} \\
$n_0$ (fm$^{-3}$) & $0.16$ & density unit \\
$Y_L$ & $0.1$, $0.2$, $0.3$ & \\
$T$ (MeV) & $5$--$20$ & \\
$n_B/n_0$ & $1.5$--$4$ & \\
\end{tabular}
\end{ruledtabular}
\end{table}

\subsection{Inverse mean free path}
For a neutrino of energy $E$ and momentum $\mathbf k$, the inverse mean free path for the reaction $\nu(k)+B_i(P_i)\to B_f(P_f)+l(q)$ is
\begin{align}
\frac{1}{\lambda(E)}={}&\frac{1}{2E}\int\frac{d^3p_i}{(2\pi)^32E_i^*}\frac{d^3p_f}{(2\pi)^32E_f^*}\frac{d^3q}{(2\pi)^32E_q}\nonumber\\
&\times(2\pi)^4\delta^4(k+P_i-P_f-q)\,\overline{\sum}|\mathcal M|^2\,F ,
\label{eq:lambda}
\end{align}
with the statistical factor
\begin{equation}
F=f_i(E_i)\,[1-f_f(E_f)]\,[1-f_l(E_q)] .
\label{eq:F}
\end{equation}
Here $E_b=E_b^*+U_b$ is the single-particle energy, $E_b^*=(p_b^2+m_b^{*2})^{1/2}$ is the kinetic energy with the Dirac effective mass $m_b^*$, and $U_b$ is the vector self-energy. The spinors and the matrix element depend on the kinetic four-momenta $P_b=(E_b^*,\mathbf p_b)$, while energy conservation involves the total single-particle energies~\cite{RobertsReddy2017,Kumar2026P1}. The spin sum runs over all baryon and lepton spin states, and the occupation numbers are Fermi--Dirac distributions in the total energies. The outgoing lepton is blocked by $1-f_e$ for electrons, $1-f_{e^+}$ for positrons, and $1-f_\nu$ for scattered neutrinos.

We integrate over the final baryon with the spatial delta function and over the magnitude of the outgoing lepton momentum with the energy delta function. Energy conservation then requires a root of
\begin{equation}
h(q)=U_f+\Big[\big|\mathbf k+\mathbf p_i-q\hat{\mathbf q}\big|^2+m_f^{*2}\Big]^{1/2}+\big(q^2+m_l^2\big)^{1/2}-E-E_i ,
\label{eq:h}
\end{equation}
where $\hat{\mathbf q}$ is the lepton direction. The function $h$ is convex in $q$, so it has at most two roots. With $\mathbf k$ along the $z$ axis and the initial baryon in the $x$--$z$ plane, the inverse mean free path becomes
\begin{align}
\frac{1}{\lambda(E)}={}&\frac{1}{2E}\,\frac{1}{8(2\pi)^4}\int dE_i\,d\cos\theta_i\,d\Omega_q \nonumber\\
&\times\sum_{\text{roots}}\frac{p_i\,q^2\,\overline{\sum}|\mathcal M|^2\,F}{E_q\,E_f^*\,|h'(q)|} ,
\label{eq:lambda4}
\end{align}
with
\begin{equation}
h'(q)=\frac{q-(\mathbf k+\mathbf p_i)\cdot\hat{\mathbf q}}{E_f^*}+\frac{q}{E_q} .
\label{eq:hp}
\end{equation}
The derivation and the root finding are given in Appendix~\ref{app:ps}. We evaluate the four-dimensional integral by quasi-Monte Carlo sampling with scrambled Sobol sequences, as implemented in SciPy~\cite{SciPy2020} (\texttt{scipy.stats.qmc}, version 1.18).

\subsection{Three-body antineutrino absorption}
\label{sec:id}
For the reaction $\bar\nu_e(k)+e^-(p_e)+H(P_H)\to L(P_L)$ the inverse mean free path is
\begin{align}
\frac{1}{\lambda(E)}={}&\frac{1}{2E}\int\frac{d^3p_e}{(2\pi)^32E_e}\frac{d^3p_H}{(2\pi)^32E_H^*}\frac{d^3p_L}{(2\pi)^32E_L^*}\nonumber\\
&\times(2\pi)^4\delta^4(k+p_e+P_H-P_L)\,\overline{\sum}|\mathcal M|^2\,F_3 ,
\label{eq:lamID}
\end{align}
with
\begin{equation}
F_3=f_e(E_e)\,f_H(E_H)\,[1-f_L(E_L)] .
\label{eq:F3}
\end{equation}
The spatial delta function sets $\mathbf p_L=\mathbf Q+\mathbf p_H$ with $\mathbf Q=\mathbf k+\mathbf p_e$, and the energy delta function fixes the angle $\alpha$ between $\mathbf p_H$ and $\mathbf Q$,
\begin{equation}
\cos\alpha=\frac{(E+E_e+E_H-U_L)^2-m_L^{*2}-Q^2-p_H^2}{2\,Q\,p_H} .
\label{eq:cosa}
\end{equation}
The result, derived in Appendix~\ref{app:ps}, is
\begin{align}
\frac{1}{\lambda(E)}={}&\frac{1}{2E}\,\frac{1}{8(2\pi)^4}\int dE_e\,d\cos\theta_e\,dE_H\,d\phi\nonumber\\
&\times\frac{p_e\,\overline{\sum}|\mathcal M|^2\,F_3}{Q} ,
\label{eq:lamID4}
\end{align}
where $\theta_e$ is the angle between $\mathbf p_e$ and $\mathbf k$, $\phi$ is the azimuth of $\mathbf p_H$ about $\mathbf Q$, and the integrand vanishes unless $|\cos\alpha|\leq1$. The squared matrix element is that of the decay $L\to H+e^-+\bar\nu_e$ (Sec.~\ref{sec:me}). We evaluate Eq.~\eqref{eq:lamID4} with the same quasi-Monte Carlo method.

The kinematics of the nucleonic and $\Delta$ reactions differ. For $\bar\nu_e+e^-+p\to n$ the final neutron must lie above its Fermi surface while the momenta of the initial proton and electron, which are near their Fermi surfaces, must add up to that of the neutron; in matter below the direct Urca threshold this is possible only in the thermal tails, as for the direct Urca process itself. For $\bar\nu_e+e^-+n\to\Delta^-$ the final $\Delta^-$ has an energy close to $\mu_n+\mu_e$, far above its own Fermi surface, so it is not Pauli blocked. The reaction is then open without thermal suppression whenever
\begin{equation}
\mu_n+\mu_e>M^*_{\Delta^-}+U_{\Delta^-} ,
\label{eq:thr}
\end{equation}
the in-medium rest energy of the $\Delta^-$; all chemical potentials include the rest masses. Energy conservation alone is not sufficient: the momentum of the $\Delta^-$ at this energy must also lie between $|p_{Fn}-p_{Fe}|$ and $p_{Fn}+p_{Fe}$, so that it can be built from a neutron and an electron at their Fermi surfaces. At $T=10$~MeV, $2$--$3n_0$ and both lepton fractions we find $p_{\Delta^-}=250$--530~MeV, inside these limits in every case. Since $\mu_{\Delta^-}=\mu_n+\mu_e-\mu_{\nu_e}$ in equilibrium, the $\Delta^-$ population is controlled by a chemical potential that is lower by $\mu_{\nu_e}$, so Eq.~\eqref{eq:thr} can hold where the $\Delta^-$ fraction is negligible. We call the difference of the two sides of Eq.~\eqref{eq:thr} the threshold margin. In vacuum the reaction would be resonant pion production, $\bar\nu_e+e^-+n\to\Delta^-\to n+\pi^-$. In matter the $\Delta^-$ near its Fermi surface is long lived (Sec.~\ref{sec:mb}), but the $\Delta^-$ produced here is far above it, where our models do not determine its width; we treat it as a quasiparticle and test the effect of a width in Sec.~\ref{sec:mb}. The direct process $\bar\nu_e+e^-\to\pi^-$ is suppressed by the electron mass, since its squared amplitude is proportional to the square of the charged-lepton mass~\cite{ForeReddy2020}; the reaction considered here involves a nucleon and the $N\Delta$ axial transition, and has no such suppression.

\subsection{Matrix elements and crossing}
\label{sec:me}
The charged- and neutral-current interactions are
\begin{align}
\mathcal L_{CC}&=-\frac{G_F\cos\theta_C}{\sqrt2}\,\bar e\gamma_\mu(1-\gamma_5)\nu_e\,J^\mu_W+\text{h.c.}, \nonumber\\
\mathcal L_{NC}&=-\frac{G_F}{\sqrt2}\,\bar\nu\gamma_\mu(1-\gamma_5)\nu\,J^\mu_Z ,
\label{eq:L}
\end{align}
with the baryonic currents of Table~\ref{tab:channels}; for nucleons this is the normalization of Ref.~\cite{RPL1998}, with our $c_V$ and $c_A$ equal to half of theirs. For every vertex the spin-summed squared matrix element takes the form
\begin{equation}
\overline{\sum}|\mathcal M|^2=32\,G^2\,|\mathcal I|^2\,K\big(P_1,P_2,\ell,\kappa\big) ,
\label{eq:MK}
\end{equation}
with $G=G_F\cos\theta_C$ for charged currents and $G=G_F$ for neutral currents. The function $K$ depends on the kinetic four-momenta $P_1$ of the lower-charge baryon and $P_2$ of the higher-charge baryon, on a charged-lepton-like momentum $\ell$ and on a neutrino-like momentum $\kappa$. For a Dirac baryon with the vertex $\gamma^\mu(c_V-c_A\gamma_5)$,
\begin{align}
K_{VA}={}&(c_V+c_A)^2(P_2\!\cdot\!\ell)(P_1\!\cdot\!\kappa)+(c_V-c_A)^2(P_1\!\cdot\!\ell)(P_2\!\cdot\!\kappa)\nonumber\\
&-(c_V^2-c_A^2)\,m_1^*m_2^*\,\ell\!\cdot\!\kappa ,
\label{eq:KVA}
\end{align}
which for $c_V=1$, $c_A=g_A$ is the familiar nucleonic result, Eq.~(7) of Ref.~\cite{RPL1998} with $M^2\to m_1^*m_2^*$, evaluated with the kinetic momenta as in Ref.~\cite{RobertsReddy2017}. For the $N\to\Delta$ transitions we use the general weak vertex~\cite{HNV2007}
\begin{align}
\Gamma^{\sigma\mu}={}&\Big[\frac{C_3^V}{M}\big(g^{\sigma\mu}\slashed{q}-q^\sigma\gamma^\mu\big)\nonumber\\
&+\frac{C_4^V}{M^2}\big(g^{\sigma\mu}q\!\cdot\!P_\Delta-q^\sigma P_\Delta^\mu\big)\nonumber\\
&+\frac{C_5^V}{M^2}\big(g^{\sigma\mu}q\!\cdot\!P_N-q^\sigma P_N^\mu\big)\Big]\gamma_5\nonumber\\
&+\frac{C_4^A}{M^2}\big(g^{\sigma\mu}q\!\cdot\!P_\Delta-q^\sigma P_\Delta^\mu\big)+C_5^A\,g^{\sigma\mu} ,
\label{eq:GND}
\end{align}
with $\sigma$ the Rarita--Schwinger index, $q=P_\Delta-P_N$ and $M$ the nucleon mass; the matrix element is $\bar u_\sigma(P_\Delta)\Gamma^{\sigma\mu}u(P_N)$ for $N\to\Delta$ and its complex conjugate for $\Delta\to N$~\cite{HNV2007}. The vector form factors $C_3^V=2.13$, $C_4^V=-1.51$ and $C_5^V=0.48$ at $q^2=0$ follow from conserved-vector-current and electroproduction data, and the axial ones are $C_5^A(0)=1.2$, $C_4^A=-C_5^A/4$ and $C_3^A=0$, all with the $q^2$ dependence of Ref.~\cite{HNV2007} (vector mass 0.84~GeV, axial mass 1.05~GeV). We omit the pion-pole term $C_6^A=C_5^AM^2/(m_\pi^2-q^2)$. Contracted with the lepton current it is proportional to the electron mass, but in the three-body kinematics $q^2$ crosses $m_\pi^2$, where the bare pole is not integrable; there it describes real pion production, $e^-+\bar\nu_e\to\pi^-$, whose squared amplitude is proportional to $m_e^2$~\cite{ForeReddy2020}. With the in-medium $\pi^-$ propagator of Sec.~\ref{sec:mb} the pole is regulated, and a Cauchy--Schwarz bound on the spin sums, integrated over the three-body phase space, limits the change of the three-body opacity from this term to 1.5\% at $T=20$~MeV for $g'=0.5$--0.7. Reference~\cite{Kumar2026P1} kept only the $C_5^A$ term. We evaluate $K$ for the vertex of Eq.~\eqref{eq:GND} by explicit summation over Dirac and Rarita--Schwinger spinors at each integration point, with the kinetic momenta and effective masses in the Dirac structure and the form factors evaluated at the squared four-momentum of the lepton pair, $(k\pm p_e)^2$, which is the momentum carried by the $W$; in the $C_5^A$-only limit it reduces to the analytic form given in Appendix~\ref{app:K}. The $\Delta$ is an external on-shell quasiparticle in all our reactions, never an exchanged one. For external on-shell $\Delta$s the conventional couplings are equivalent to the consistent couplings that decouple the spin-$\tfrac12$ components of the Rarita--Schwinger field: Ref.~\cite{Pascalutsa:1999zz} constructed consistent $\pi N\Delta$ and $\gamma N\Delta$ couplings that coincide with the conventional ones when contracted with an on-shell Rarita--Schwinger spinor (they differ for $\Delta$ exchange), and Ref.~\cite{Krebs:2008zb} showed by a field redefinition that the two differ on shell only by contact terms. We assume that the same holds for the weak vertex of Eq.~\eqref{eq:GND}. For $\Delta\to\Delta'$, $K$ is the spin-summed polynomial of Ref.~\cite{Kumar2026P1} with the vector and axial couplings separated,
\begin{equation}
K_{\Delta\Delta'}=\frac{1}{64}\Big(c_V^2\,t_0+c_Vc_A\,t_1+c_A^2\,t_2\Big) ,
\label{eq:KDD}
\end{equation}
where $t_0$, $t_1$ and $t_2$ are the parts of zeroth, first and second order in the axial coupling. The polynomial and the $C_5^A$-only limit of the $\Delta N$ function are given in Appendix~\ref{app:K}.

The decay matrix elements are crossed to the reactions of Eqs.~\eqref{eq:cc} and~\eqref{eq:nc}. Every term of $K$ is bilinear in the two lepton momenta, so reversing both lepton lines leaves $K$ unchanged, while reversing one lepton line and one baryon line exchanges the roles of the two baryons. The resulting assignments are
\begin{align}
\nu_e+L\to H+e^-:&\quad P_1=P_i,\ P_2=P_f,\ \ell=q,\ \kappa=k, \nonumber\\
\bar\nu_e+H\to L+e^+:&\quad P_1=P_f,\ P_2=P_i,\ \ell=q,\ \kappa=k, \nonumber\\
\nu+B\to\nu+B:&\quad P_1=P_i,\ P_2=P_f,\ \ell=q,\ \kappa=k, \nonumber\\
\bar\nu+B\to\bar\nu+B:&\quad P_1=P_i,\ P_2=P_f,\ \ell=k,\ \kappa=q , \nonumber\\
\bar\nu_e+e^-+H\to L:&\quad P_1=P_L,\ P_2=P_H,\ \ell=p_e,\ \kappa=k .
\label{eq:cross}
\end{align}
The last line is the decay matrix element itself, with all momenta reversed. We verified each of the resulting ten crossed forms (three vertex types, crossed to the four two-body reactions where they apply) against explicit Dirac and Rarita--Schwinger spinor sums, and the three-body opacity against an independent decay-rate code (Sec.~\ref{sec:valid}).

\subsection{Static limit for spin-$\tfrac32$ targets}
\label{sec:static}
For a slow, nondegenerate target and a neutrino energy small compared to the target mass, neutral-current scattering reduces to the vector (Fermi) and axial (Gamow--Teller) strengths of the target. For the nucleon the cross section is~\cite{RPL1998}
\begin{equation}
\sigma_N=\frac{G_F^2E^2}{\pi}\big(c_V^2+3c_A^2\big) .
\label{eq:sigN}
\end{equation}
For the $\Delta$ the axial current of Table~\ref{tab:channels} acts on the Dirac spin $\mathbf s$ of the Rarita--Schwinger state only. In the rest frame the spin-$\tfrac32$ state is the coupling of $\mathbf s$ to a vector index of orbital-like spin $l=1$, and within the multiplet of total spin $\mathbf S$ the projection theorem gives
\begin{equation}
\bm\sigma_{\rm eff}=2\,\frac{\langle\mathbf S\!\cdot\!\mathbf s\rangle}{S(S+1)}\,\mathbf S=\frac23\,\mathbf S ,
\label{eq:proj}
\end{equation}
where we used
\begin{equation}
\langle\mathbf S\!\cdot\!\mathbf s\rangle=\tfrac12\big[S(S+1)+s(s+1)-l(l+1)\big]=\tfrac54 .
\label{eq:Ss}
\end{equation}
The Gamow--Teller strength summed over final and averaged over initial spin states is then
\begin{equation}
\frac{1}{2S+1}\sum_{m,m'}\sum_{a}\big|\langle m'|\sigma^a_{\rm eff}|m\rangle\big|^2=\frac49\,S(S+1)=\frac53 ,
\label{eq:GT32}
\end{equation}
compared to three for the nucleon. We checked Eq.~\eqref{eq:GT32} by constructing the spin-$\tfrac32$ states from Clebsch--Gordan coefficients and summing the matrix elements of $\bm\sigma$ explicitly. The cross section on a $\Delta$ is therefore
\begin{equation}
\sigma_\Delta=\frac{G_F^2E^2}{\pi}\Big(c_V^2+\frac53\,c_A^2\Big) .
\label{eq:sigD}
\end{equation}
The same factor $5/3$ appears in the static limit of the charged-current $\Delta\to\Delta'$ matrix element. Equation~\eqref{eq:sigD} shows that, for a given coupling $g_{A\Delta}$ in the vertex of Table~\ref{tab:channels}, the higher spin of the $\Delta$ does not increase its axial strength per particle; it reduces it, because only the Dirac spin carries the axial current. A different normalization of the $\Delta$ axial coupling, for example one defined through the total spin, would absorb part of this factor into the coupling. Reference~\cite{Chen2007} attributed the dominance of the $\Delta^-$ in neutrino scattering partly to the high spin of the $\Delta$, from a comparison of the response of a fictitious spin-$\tfrac32$ particle with that of the neutron at equal effective mass and chemical potential, where the fourfold spin degeneracy doubles the density of states. Per particle, at least in the static limit, the spin does not raise the vector response and lowers the axial one. What distinguishes the $\Delta^-$ is its vector charge: $c_V^2$ is about four times its neutron value, and it is fixed by the conservation of the vector current, independently of the poorly known $\Delta$ axial coupling. It equals the vector charge of the $\Sigma^-$ and $\Xi^-$ hyperons~\cite{RPL1998}, whose contribution to scattering in hyperonic matter is known to be significant~\cite{Reddy:1996tw}. With $g_{A\Delta}=0.6\,g_A$ the vector term supplies one third of $\sigma_{\Delta^-}$, and $\sigma_{\Delta^-}$ is about 2.2 times the cross section on a neutron. In degenerate matter a further difference arises from Pauli blocking: the $\Delta^-$, distributed over four spin states and less abundant than the neutrons, have a much smaller Fermi energy, so a larger fraction of them can scatter.

\subsection{Diffusion mean free path}
\label{sec:diff}
We characterize transport by the diffusion coefficients of the equilibrium diffusion approximation~\cite{Pons:1998mm}. There the energy-dependent diffusion coefficient is the inverse of the sum of emissivity, absorptivity and transport scattering opacity. By detailed balance, emission and absorption combine to the absorption opacity corrected for stimulated absorption, $\kappa_a^*=\kappa_a/(1-f_\nu)$, with $\kappa_a=1/\lambda_a$ summed over the charged-current channels, including the three-body absorption for antineutrinos. Scattering enters through the transport opacity
\begin{equation}
\kappa_{s}^{\rm tr}(E)=\int d\Omega_q\,(1-\cos\theta_{kq})\,\frac{d\kappa_s}{d\Omega_q} ,
\label{eq:tr}
\end{equation}
where $d\kappa_s/d\Omega_q$ is evaluated with the factor $1-f_\nu(E')\big[1-e^{(E'-E)/T}\big]$ for the scattered neutrino of energy $E'$ in place of the Pauli-blocking factor $1-f_\nu(E')$. This factor combines out-scattering with blocking and in-scattering; it is the statistical factor of the isotropic term of Eq.~(25) of Ref.~\cite{Pons:1998mm}. We apply it also to the anisotropic ($\cos\theta$) term, which in that equation carries the ratio of the first angular moments of the distribution at $E'$ and $E$ and a different statistical factor. For nearly elastic scattering on baryons the dependence on the anisotropic moment can be ignored~\cite{Pons:1998mm}, and the approximation is exact for isoenergetic scattering, for which the factor equals one and Eq.~\eqref{eq:tr} reduces to the familiar transport opacity without blocking [Eq.~(26) of Ref.~\cite{Pons:1998mm}]. We evaluate it with the full kinematics, including the energy transfer to the baryon. The diffusion mean free path is
\begin{equation}
\lambda_R=\frac{\int dE\,E^2 f_\nu(1-f_\nu)\,\big[\kappa_a^*(E)+\kappa_s^{\rm tr}(E)\big]^{-1}}{\int dE\,E^2 f_\nu(1-f_\nu)} ,
\label{eq:lamR}
\end{equation}
where $f_\nu$ is the equilibrium distribution of the species considered, at chemical potential $\mu_{\nu_e}$ for neutrinos and $-\mu_{\nu_e}$ for antineutrinos. The weight $E^2f_\nu(1-f_\nu)$ is proportional to the derivative of the number density with respect to the chemical potential, so $\lambda_R$ controls the diffusion of lepton number: it is the coefficient $D_2$ of Ref.~\cite{Pons:1998mm}, normalized by the same integral without the opacity, which multiplies the gradient of the neutrino degeneracy in the lepton-number flux. Because $\lambda_R$ is a harmonic mean of the opacity, it is dominated by the energies where the opacity is smallest within the window of the weight.

\subsection{Many-body corrections}
\label{sec:mb}
The quasiparticle opacities above neglect the collisional width of the baryons and the correlations among them. In nucleonic matter the collisional width~\cite{RobertsReddyShen2012} and correlations~\cite{Shin:2023sei} modify the charged-current opacities. We estimate both.

\textit{Collisional broadening, charged current.} We use the nucleon width approximation (NWA)~\cite{AHZ2024nwa}, as in Ref.~\cite{Kumar2026P1}: the effective masses of the initial and final baryons are smeared independently with Breit--Wigner distributions of full width $W=T^2/T_W$ with $T_W=5$~MeV. This width, $\gamma=aT^2$ with $a\approx0.2$~MeV$^{-1}$, was obtained from a finite-temperature Brueckner calculation for pure neutron matter at one to two times saturation density, where the $T^2$ law holds up to somewhat below 30~MeV~\cite{SedrakianDieperink2000}, and has only a weak density dependence~\cite{SedrakianDieperink2000,AHZ2024nwa}; at 3--$4n_0$ it is an extrapolation, and we use the same width for the $\Delta$s, which is an assumption. At $T\lesssim10$~MeV this width is no larger than the intrinsic isospin breaking of the charged current in dense matter~\cite{AHZ2024nwa}; at 20~MeV it is 80~MeV, and the approximation is less well founded. For the thermal $\Delta$ population the $\Delta\to N\pi$ decay is expected to be Pauli blocked~\cite{RadutaOertelSedrakian2020}, as found in neutrino-transparent matter up to 10~MeV~\cite{Kumar2026P1}, and we take the $\Delta$ width to be the collisional one. This argument does not cover the $\Delta^-$ produced in the three-body reaction, which lies far above its Fermi surface. We computed its $\Delta^-\to n\pi^-$ width with the $\pi N\Delta$ coupling $f^*=2.14$ of Ref.~\cite{HNV2007}, explicit Rarita--Schwinger spinors, the mean-field energies and the Pauli blocking of the final neutron, and vacuum pions; in vacuum the same calculation reproduces the relativistic $\Delta$ width. For the $\Delta^-$ produced by antineutrinos of energy $T$ to $3T$ at 2--$3n_0$, with momenta of 250--590~MeV, the width with vacuum pions is 6--30~MeV, much smaller than in vacuum because the neutron must be placed above its Fermi surface.

We then replaced the vacuum pion by the in-medium $\pi^-$ propagator $D=[\omega^2-k^2-m_\pi^2-\Pi(\omega,k)]^{-1}$ and computed
\begin{align}
\Gamma={}&\frac{1}{2E^*_\Delta}\int_0^\infty\frac{k\,dk}{4\pi p_\Delta}\int d\omega\,S(\omega,k)\,\overline{|\mathcal M|^2}\,|V|^2\nonumber\\
&\times(1-f_n)\,\big[1+f_B(\omega-\mu_\pi)\big],
\label{eq:Gmed}
\end{align}
with $S=-\mathrm{Im}\,D/\pi$, $\mu_\pi=\mu_n-\mu_p=\mu_e-\mu_{\nu_e}$ and $f_B$ the Bose function. The integral runs over the pion branch and over the spin-isospin continuum, so it includes $\Delta N\to NN$ by pion exchange. The self-energy is the sum of the leading-order Weinberg--Tomozawa term $\Pi_s=\omega(n_n-n_p)/2f_\pi^2$, which dominates the $\pi^-$ mass at zero momentum in neutron-rich matter up to about saturation density~\cite{ForeKaiserReddyWarrington2024}, and the $p$-wave nucleon-hole and $\Delta$-hole polarization $\chi$ with Landau--Migdal correlations, $\Pi_p=-k^2F^2\chi/(1+g'\chi)$, the single-$g'$ limit of the form in Ref.~\cite{HelgessonRandrup1995}. We evaluated $\chi$ as a Lindhard function at finite temperature with the mean-field energies of $n$, $p$, $\Delta^-$ and $\Delta^0$, the $\pi NN$ coupling $f^2/4\pi=0.08$, $f^*=2.14$, a monopole form factor $F$ with cutoff 1~GeV and the vertex correction $V=1/(1+g'\chi)$. The $\Delta$ occupations enter in the form $f_N-f_\Delta$, which makes $\mathrm{Im}\,\Pi$ change sign at $\omega=\mu_\pi$ as detailed balance requires; we checked this at every state, together with the static Lindhard and dilute $\Delta$-hole limits and a Kramers--Kronig relation, and with $\Pi=0$ Eq.~\eqref{eq:Gmed} reproduces the vacuum-pion width to 0.3\%. We give the resulting widths, and the stability of the matter against pion condensation, in Sec.~\ref{sec:pion}.

 In the three-body channel we therefore give the final $\Delta$ a Breit--Wigner width $W=T^2/T_W$, which is comparable to or larger than this estimate, and, as a large-width test, the vacuum Breit--Wigner width of 117~MeV~\cite{PDG2024}, keeping the initial baryon on shell.

\textit{Collisional broadening, neutral current.} Giving the baryons a width without the corresponding vertex corrections violates the Ward identity of the conserved vector current, so the NWA is not appropriate for neutral-current scattering~\cite{AHZ2024nwa}; we also found that off-shell Lorentzian spectral functions in the occupation numbers do not conserve the baryon density. We therefore broaden the baryonic response in the energy transfer $q_0=E-E_q$. The response obeys detailed balance~\cite{BurrowsSawyer1998},
\begin{equation}
S(-q_0)=e^{-q_0/T}\,S(q_0) ,
\label{eq:db}
\end{equation}
so the function $\tilde S(q_0)=e^{q_0/2T}S(q_0)$ is even, and a convolution of $\tilde S$ with an even kernel is again even. The broadened response
\begin{equation}
S_\Gamma(q_0)=\frac{\int d\varepsilon\,g_\sigma(\varepsilon)\,e^{-\varepsilon/2T}\,S(q_0-\varepsilon)}{\int d\varepsilon\,g_\sigma(\varepsilon)\,e^{-\varepsilon/2T}}
\label{eq:SG}
\end{equation}
therefore satisfies Eq.~\eqref{eq:db} exactly. We take $g_\sigma$ to be a Gaussian with full width at half maximum $T^2/T_W$ and keep the baryon occupations on shell, so the density is unchanged. Details are given in Appendix~\ref{app:width}.

\textit{Long-wavelength correlations.} In the long-wavelength, static limit the response of the matter is determined by its thermodynamics, which makes the correlation correction consistent with the equation of state~\cite{RPLP1999}. In the mean-field approximation the static density response of the multicomponent matter at $q\to0$ is the random-phase sum of meson exchanges, which equals the thermodynamic susceptibility matrix,
\begin{equation}
\chi_{ij}=\frac{\partial n_i}{\partial\mu_j}\bigg|_T=\Big[(1-\chi^0V)^{-1}\chi^0\Big]_{ij} .
\label{eq:chi}
\end{equation}
Here all species chemical potentials are varied independently, $\chi^0_{ij}=\delta_{ij}\,\partial n_i/\partial\nu_i$ is the response at frozen meson fields with $\nu_i=\mu_i-U_i$, and $V$ is the meson-exchange interaction including the rearrangement terms of the density-dependent couplings. The weak vector structure factor is reduced by
\begin{equation}
R_V=\frac{c^{\,T}\chi\,c}{c^{\,T}\chi^0c} ,
\label{eq:RV}
\end{equation}
where $c$ is the vector of weak vector charges. We apply $R_V$ to the vector part of the neutral-current opacity, obtained by setting $c_A=0$. The axial channel is not corrected, because the mean-field models contain no spin-dependent residual interaction; in $\sigma\omega\rho$ models the RPA indeed modifies only the vector response~\cite{RPLP1999}. Coulomb correlations are neglected. Since the static $q\to0$ limit overestimates the screening at the momentum and energy transfers of thermal neutrinos, this is an upper bound on the effect of correlations in the vector channel.

\section{Validation}
\label{sec:valid}
We checked the calculation in three independent ways. First, the charged-current opacity is related by detailed balance to the electron-capture rate, which we computed with the independent rate code of Ref.~\cite{Kumar2026P1}, a full phase-space calculation in the approach of Refs.~\cite{AHZ2024nwa,Tambe2026}; the code uses a different parametrization of the phase space, so this tests the kinematics, the root finding and the crossing of the matrix elements for the neutrino reactions. The antineutrino charged-current forms, which have no counterpart in that code, are tested by the second check. The three-body absorption of Eq.~\eqref{eq:id} is related by detailed balance to the rate of the decay $L\to H+e^-+\bar\nu_e$, which the same code computes with its own phase-space parametrization; this tests Eqs.~\eqref{eq:lamID4} and~\eqref{eq:cosa} and the matrix elements of the three-body channel. Second, every spin-summed function $K$, in each of its crossed forms, was compared with explicit sums over Dirac and Rarita--Schwinger spinors at random relativistic kinematics; for the full $N\Delta$ vertex the comparison is with a second, independent implementation of the spinor sums. We also checked that the vector part of Eq.~\eqref{eq:GND} is conserved and that the three-body reduction of Eq.~\eqref{eq:lamID4} agrees with a brute-force integration over both momenta in a test case. Third, in a dilute, nondegenerate gas the opacities must reduce to the static limits of Eqs.~\eqref{eq:sigN} and~\eqref{eq:sigD}, and the broadened neutral-current opacity must reproduce the analytic shift of the thermal average. In addition, the susceptibility matrix of Eq.~\eqref{eq:chi} must be symmetric, which tests the derivatives of the density-dependent couplings, and the Gamow--Teller strength of Eq.~\eqref{eq:GT32} was checked by an explicit sum over spin-$\tfrac32$ states. Detailed balance with the independent code holds to 0.99--1.01 for the two-body reactions and to 0.98--1.003 for the three-body reactions (with the $C_5^A$-only vertex, which that code contains); the spinor sums agree to $10^{-10}$ or better; the three-body reduction agrees with the brute-force integral to $1.004\pm0.008$; the dilute-gas limits are reproduced to 1--3\%; the transport factor of Eq.~\eqref{eq:tr} reproduces the unblocked transport opacity in the elastic limit (to 0.1\% for a target mass of $3\times10^6$~MeV, with a deviation that falls as the inverse mass); and the spin-$\tfrac32$ sum gives exactly $5/3$. The quasi-Monte Carlo and energy-grid errors are discussed in Appendix~\ref{app:conv}. As an external benchmark we compared with the neutral-current mean free paths of Ref.~\cite{Chen2007} (Appendix~\ref{app:chen}): the $\Delta$-induced reduction agrees, while the absolute values differ by a factor of \chenFactor.

\section{Results}
\label{sec:results}
\subsection{Composition and the neutrino chemical potential}
\begin{figure*}[tbp]
\includegraphics[width=\textwidth]{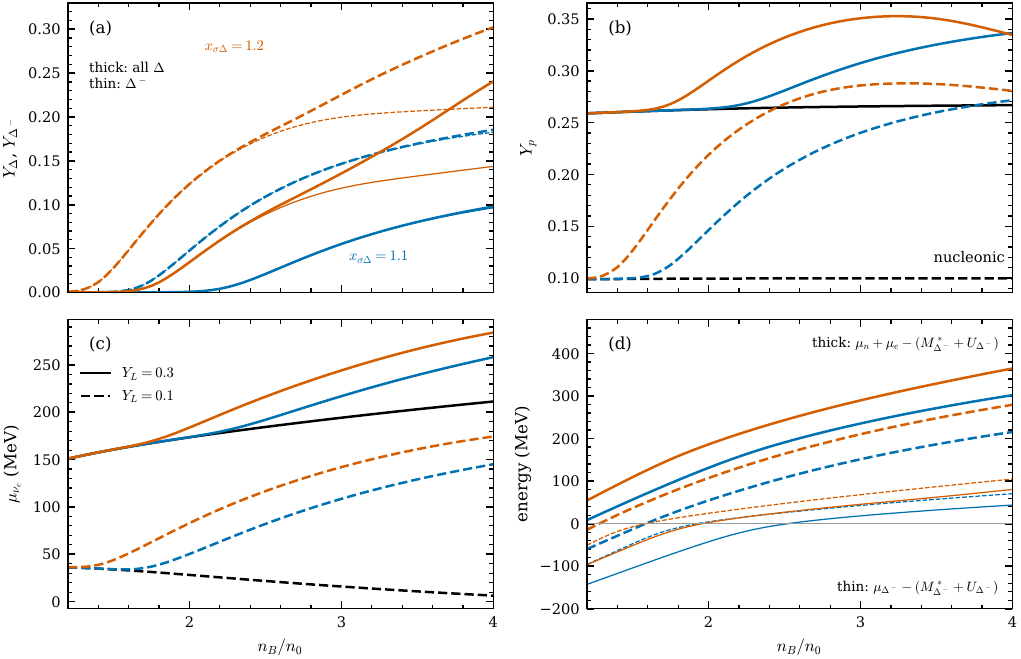}
\caption{Composition of neutrino-trapped matter at $T=10$~MeV as a function of baryon density, for DDME2 without $\Delta$s (black) and with $\Delta$s for $x_{\sigma\Delta}=1.1$ (blue) and $1.2$ (orange), at lepton fractions $Y_L=0.3$ (solid) and $0.1$ (dashed). (a) Total $\Delta$ fraction (thick) and $\Delta^-$ fraction (thin); (b) proton fraction; (c) electron-neutrino chemical potential; (d) threshold margin of the three-body reaction $\bar\nu_e+e^-+n\to\Delta^-$, $\mu_n+\mu_e-(M^*_{\Delta^-}+U_{\Delta^-})$ (thick), and $\mu_{\Delta^-}-(M^*_{\Delta^-}+U_{\Delta^-})$ (thin), which differ by $\mu_{\nu_e}$. The $\Delta^-$ displaces electrons, which raises the neutrino chemical potential and the proton fraction; the antineutrino channel opens where the margin is positive, at lower density than the $\Delta^-$ population.}
\label{fig:comp}
\end{figure*}

Figure~\ref{fig:comp} shows how the $\Delta$s change the composition of neutrino-trapped matter. In DDME2 at $T=10$~MeV and $Y_L=0.3$ the $\Delta^-$ appears (fraction above 0.01) at $1.8n_0$ for $x_{\sigma\Delta}=1.2$ and at $2.35n_0$ for $x_{\sigma\Delta}=1.1$; at $Y_L=0.1$ it appears about $0.5n_0$ earlier. The total $\Delta$ fraction at $3$--$4n_0$ is 0.05--0.1 ($Y_L=0.3$) and 0.15--0.19 ($Y_L=0.1$) for $x_{\sigma\Delta}=1.1$, and 0.14--0.24 and 0.23--0.30 for $x_{\sigma\Delta}=1.2$ [Fig.~\ref{fig:comp}(a)]; below about $3n_0$ it consists almost entirely of $\Delta^-$ (thin lines). For $x_{\sigma\Delta}=1.2$ the $\Delta^0$ appears near $2.8n_0$ and the $\Delta^+$ at 3.7--$4n_0$, the same ordering of thresholds as found for hot matter in other models and couplings~\cite{Malfatti2019,RadutaOertelSedrakian2020}.

The $\Delta^-$ takes over part of the negative charge from the electrons. Since the lepton fraction is fixed, the electrons it displaces are replaced by neutrinos, so the net neutrino fraction and the neutrino chemical potential rise. Charge neutrality then requires more protons, and the proton fraction rises [Fig.~\ref{fig:comp}(b)]. Negatively charged hyperons act in the same way in proto-neutron star matter, where the neutrino fraction also begins to increase with density once they appear~\cite{Prakash:1996xs,Pons:1998mm}. Reference~\cite{Chen2007} reports instead that in their model with $\Delta$s the neutrino abundance does not increase with density. In their model, at $T=30$~MeV and $Y_L=0.4$, the neutrino fraction decreases smoothly with density, so that the electron fraction rises~\cite{Chen2007}, and, as read from their Fig.~2, all four charge states of the $\Delta$ appear between about 2 and $3n_0$; we infer that the positively charged $\Delta$s compensate the charge of the $\Delta^-$, so that it does not displace electrons. In our models the $\Delta^-$ appears about $1n_0$ before the $\Delta^0$, and the positively charged $\Delta$s appear only near $4n_0$, if at all. The effect on the neutrino chemical potential is largest at low lepton fraction [Fig.~\ref{fig:comp}(c)]. In nucleonic matter at $Y_L=0.1$ the neutrino chemical potential decreases with density and nearly vanishes at $4n_0$; with $\Delta$s it increases with density to above 100~MeV. At $Y_L=0.1$ the nucleonic proton fraction stays at 0.10, just below the direct Urca threshold of cold $npe$ matter, $1/9\approx0.11$ from momentum conservation with equal proton and electron Fermi momenta (11--15\%~\cite{LPPH1991}), and the $\Delta^-$ raises it to 0.24--0.29 at $3n_0$. Both changes affect the absorption of electron neutrinos, as we discuss below. They also suppress the antineutrino population, whose chemical potential is $-\mu_{\nu_e}$.

Figure~\ref{fig:comp}(d) shows the threshold margin $\mu_n+\mu_e-(M^*_{\Delta^-}+U_{\Delta^-})$ of the three-body reaction $\bar\nu_e+e^-+n\to\Delta^-$ [Eq.~\eqref{eq:thr}], together with $\mu_{\Delta^-}-(M^*_{\Delta^-}+U_{\Delta^-})$, which controls the $\Delta^-$ population. In $\beta$ equilibrium $\mu_{\Delta^-}=\mu_n+\mu_e-\mu_{\nu_e}$, so the two differ by $\mu_{\nu_e}$. At $Y_L=0.3$ the margin is positive already at $1.2n_0$, and it is 160--175~MeV where the $\Delta^-$ fraction reaches 0.01; the antineutrino channel therefore opens about $1n_0$ below the density at which $\Delta^-$ appear. At $Y_L=0.1$, where $\mu_{\nu_e}$ is smaller, the two thresholds nearly coincide: the margin becomes positive at 1.3--$1.6n_0$ and is about 20~MeV at the $\Delta^-$ onset.

Our matter contains no muons. To test this we added muons in $\beta$ equilibrium, $\mu_n-\mu_p=\mu_e-\mu_{\nu_e}=\mu_\mu-\mu_{\nu_\mu}$, with the trapped muon lepton number set to zero, $Y_\mu+Y_{\nu_\mu}=0$, as for matter that had no muons at collapse, and alternatively with $\mu_{\nu_\mu}=0$. In trapped matter $\mu_n-\mu_p$ is only 50--140~MeV, so the muon chemical potential lies below the muon mass, and the muon fraction is at most $2\times10^{-3}$ for $Y_{L\mu}=0$ and $1.4\times10^{-2}$ for $\mu_{\nu_\mu}=0$ at the states of Fig.~\ref{fig:comp} with $T=10$ and 20~MeV and 2--$3n_0$. The threshold margin of the three-body reaction changes by less than 0.1~MeV and 0.5~MeV, respectively, and the $\Delta^-$ fraction by at most 2\% and 13\%, so muons do not affect our results.

\subsection{Energy-resolved opacities}
\label{sec:kE}
\begin{figure*}[tbp]
\includegraphics[width=\textwidth]{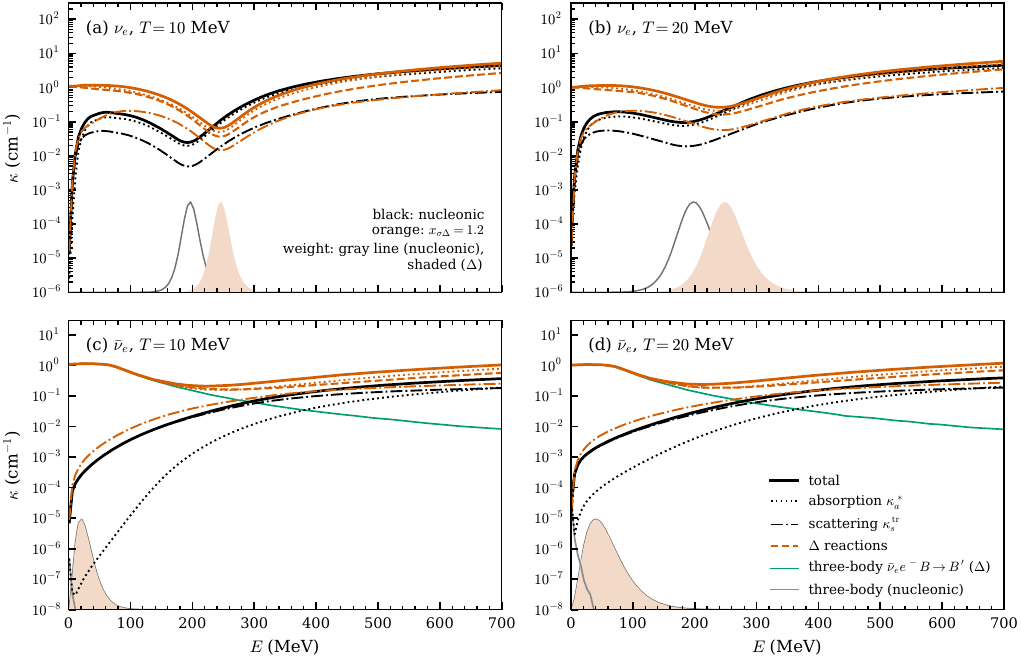}
\caption{Energy-resolved opacities at $n_B=3n_0$ and $Y_L=0.3$ for electron neutrinos (a,b) and antineutrinos (c,d), at $T=10$~MeV (a,c) and 20~MeV (b,d), in nucleonic DDME2 (black) and with $\Delta$s for $x_{\sigma\Delta}=1.2$ (orange). Solid: total opacity $\kappa_a^*+\kappa_s^{\rm tr}$; dotted: absorption corrected for stimulated absorption; dash-dotted: transport scattering opacity; dashed: contribution of all reactions involving $\Delta$s. In (c,d) the thin green and gray curves are the three-body absorption opacities with and without $\Delta$s, which are included in the absorption and total opacities. The shaded curve (with $\Delta$s) and the gray line (nucleonic) are the diffusion weight $E^2f_\nu(1-f_\nu)$ of Eq.~\eqref{eq:lamR} in arbitrary units. The absorption opacity $\kappa_a^*=\kappa_a/(1-f_\nu)$ includes the correction for stimulated absorption, which makes it large below the neutrino Fermi energy. Quasiparticle approximation. For neutrinos the $\Delta$ reactions supply about half of the opacity in the diffusion window; for antineutrinos the three-body reaction $\bar\nu_e+e^-+n\to\Delta^-$ exceeds all other reactions in the window by two to four orders of magnitude.}
\label{fig:kappaE}
\end{figure*}

Figure~\ref{fig:kappaE} shows the energy-resolved opacities at $3n_0$ and $Y_L=0.3$, at $T=10$ and 20~MeV, together with the weight of Eq.~\eqref{eq:lamR} (shaded). For electron neutrinos [Fig.~\ref{fig:kappaE}(a,b)] absorption dominates at all energies. The diffusion window lies at the minimum of the opacity, near the neutrino Fermi energy. Above it the absorption opacity rises, because the final electrons are increasingly unblocked; below it the plotted $\kappa_a^*=\kappa_a/(1-f_\nu)$ rises again, because the correction for stimulated absorption divides by the small fraction of empty neutrino states. With $\Delta$s the window moves to higher energies, since $\mu_{\nu_e}$ is larger, and the $\Delta$ reactions (dashed) supply about half the opacity in it. The largest $\Delta$ contributions are the absorption reactions $\nu_e+\Delta^-\to n+e^-$ and $\nu_e+\Delta^-\to\Delta^0+e^-$, followed by neutral-current scattering on the $\Delta^-$. The $N\to\Delta$ excitations $\nu_e+n\to\Delta^++e^-$ and $\nu_e+p\to\Delta^{++}+e^-$ contribute about one percent, because their thresholds lie well above the nucleon Fermi surfaces.

For antineutrinos [Fig.~\ref{fig:kappaE}(c,d)] the diffusion window lies at energies of a few $T$, because $\mu_{\bar\nu_e}=-\mu_{\nu_e}$ is large and negative. Charged-current absorption on protons is suppressed: the final neutron must be pushed above its Fermi surface while the positron takes energy away, and in neutron-rich matter the mean-field shift $U_n-U_p$ lowers the energy available for the reaction further~\cite{RobertsReddyShen2012}. In nucleonic matter the opacity in the window is therefore set by neutral-current scattering on neutrons and protons. The three-body absorption $\bar\nu_e+e^-+p\to n$ (gray; at 10~MeV it lies below the plotted range) exceeds two-body absorption only below about 10~MeV, and at $3n_0$ it changes $\lambda_R$ by about 1\%. At lower densities it matters more: in DDME2 it shortens $\lambda_R$ by up to 9\% at $2n_0$ and 24\% at $1.5n_0$, and in QMC-RMF3 by up to 31\% at $2n_0$, consistent with the importance of inverse neutron decay for low-energy antineutrinos found near the proto-neutron star neutrinosphere, at lower densities~\cite{Fischer2020}.

With $\Delta$s the three-body reaction $\bar\nu_e+e^-+n\to\Delta^-$ (green) changes the picture. Its opacity is 0.6--1.1~cm$^{-1}$, nearly independent of temperature and slowly decreasing with the antineutrino energy up to 100~MeV, and two to four orders of magnitude above that of all two-body reactions in the diffusion window. The energy dependence is weak because the neutron and electron Fermi seas supply most of the energy needed to produce the $\Delta^-$; the antineutrino supplies a small fraction. Two-body absorption and scattering take over only above about 300~MeV, outside the window. Among the two-body reactions, scattering on the $\Delta^-$ supplies about 40--60\% of the opacity in the window and $\bar\nu_e+\Delta^0\to\Delta^-+e^+$ adds 3--12\%, so that without the three-body channel the opacity in the window would increase by a factor of two to three. The reaction $\bar\nu_e+e^-+p\to\Delta^0$ is weaker: it adds 9\% to $\bar\nu_e+e^-+n\to\Delta^-$ at $2n_0$, $T=10$~MeV and $x_{\sigma\Delta}=1.1$, and 27\% at $3n_0$, $T=20$~MeV and $x_{\sigma\Delta}=1.2$, where the proton fraction is larger.

\subsection{Diffusion mean free paths}
\begin{figure*}[tbp]
\includegraphics[width=\textwidth]{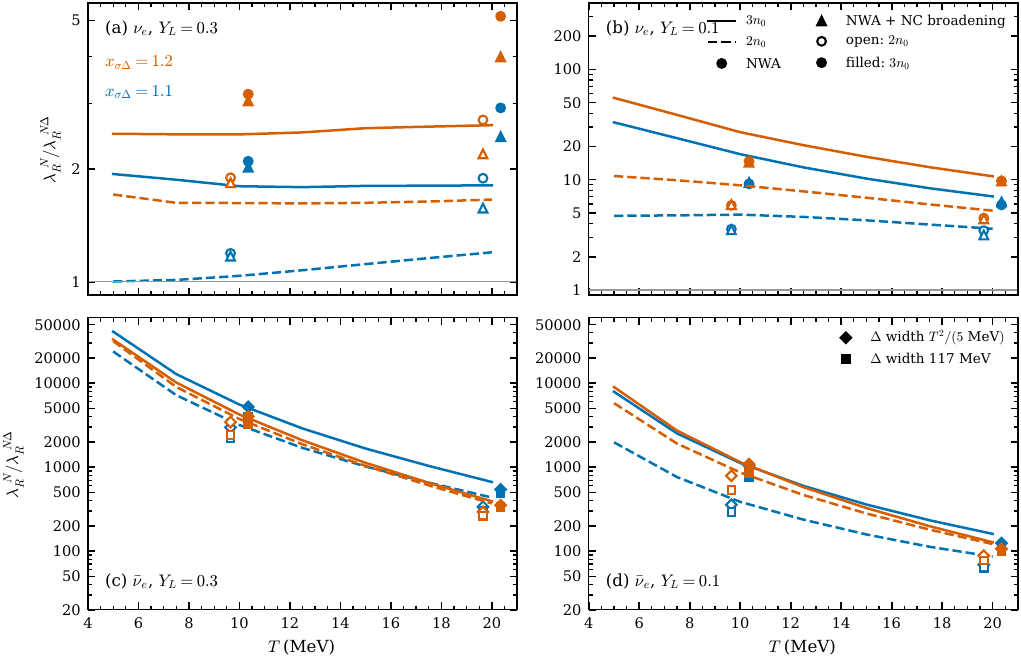}
\caption{Ratio of the diffusion mean free path in nucleonic matter, $\lambda_R^{N}$, to that in $\Delta$-admixed matter, $\lambda_R^{N\Delta}$, as a function of temperature, for electron neutrinos (a,b) and antineutrinos (c,d) at lepton fractions $Y_L=0.3$ (a,c) and $0.1$ (b,d), for DDME2 with $x_{\sigma\Delta}=1.1$ (blue) and $1.2$ (orange). Lines: quasiparticle approximation at $n_B=3n_0$ (solid) and $2n_0$ (dashed). (a,b) Circles: nucleon width approximation for all reactions; triangles: width approximation for absorption and detailed-balance-preserving energy broadening for scattering, both with $W=T^2/(5~\text{MeV})$. (c,d) Both mean free paths include three-body absorption; diamonds and squares: Breit--Wigner width $T^2/(5~\text{MeV})$ and 117~MeV of the final $\Delta$ in the three-body channel. Filled symbols: $3n_0$; open symbols: $2n_0$; the symbols are displaced by $\pm0.35$~MeV from $T=10$ and 20~MeV for legibility. The $\Delta$s shorten the mean free path in every case; for antineutrinos the three-body reaction $\bar\nu_e+e^-+n\to\Delta^-$ shortens it by two to four orders of magnitude.}
\label{fig:ratio}
\end{figure*}

Figure~\ref{fig:ratio} shows the ratio of the diffusion mean free paths without and with $\Delta$s as a function of temperature. The $\Delta$s shorten the mean free path of both species at all densities, temperatures and lepton fractions where they are present.

For electron neutrinos at $Y_L=0.3$ [Fig.~\ref{fig:ratio}(a)] the ratio reaches 1.7 at $2n_0$ and 1.8--2.6 at $3n_0$, depending on $x_{\sigma\Delta}$, and it depends little on temperature. To separate the direct contribution of the $\Delta$ reactions from the indirect one through the composition, we compare the opacity of the nucleonic reactions alone in $\Delta$-admixed matter with the total opacity of nucleonic matter. At $Y_L=0.2$ and 0.3 the two differ by $-6\%$ to $+42\%$ at 2--$4n_0$, while the $\Delta$ reactions supply 15--80\% of the opacity where the $\Delta$ fraction exceeds 0.01. The enhancement there is therefore direct: the $\Delta^-$ adds absorption channels without changing nucleonic absorption much, since in nucleonic matter at these lepton fractions the proton fraction is already above the direct Urca threshold.

At $Y_L=0.1$ [Fig.~\ref{fig:ratio}(b)] the ratio is much larger, 7--11 at 20~MeV and up to 55 at 5~MeV at $3n_0$ in the quasiparticle approximation, and it decreases with temperature. Here the indirect effect dominates. In nucleonic matter the proton fraction is just below the direct Urca threshold and $\mu_{\nu_e}$ is small, so nucleonic absorption is kinematically suppressed and the diffusing neutrinos have low energies. The $\Delta^-$ raises the proton fraction above the threshold and moves the neutrinos to energies above 100~MeV, which increases the opacity of the nucleonic reactions alone by factors of 2--15 at 2--$3n_0$, and the $\Delta$ reactions add their share on top. Hyperons also raise the neutrino chemical potential, and with it the neutrino energies, in lepton-rich matter~\cite{Reddy:1996tw}. Since a higher temperature raises both the proton fraction and the neutrino energies and so lifts the suppression of nucleonic absorption near threshold~\cite{RPL1998,Pons:1998mm}, the ratio decreases with temperature. Near threshold the quasiparticle approximation exaggerates the suppression, and collisional broadening reduces the ratio by up to a factor of two (open circles and crosses), but at $3n_0$ it remains 6--15.

At $Y_L=0.3$ collisional broadening has the opposite effect: at $3n_0$ it increases the neutrino ratio by 12--28\% at 10~MeV and by factors of 1.35--1.96 at 20~MeV. At 20~MeV the broadening lowers the opacity of nucleonic matter and raises that of $\Delta$-admixed matter. At 20~MeV the assumed width, $W=80$~MeV, is large, and this result is less certain than the quasiparticle one; in this sense the quasiparticle ratios are conservative.

For antineutrinos [Fig.~\ref{fig:ratio}(c,d)] the three-body channel makes the ratio very large. At 2--$3n_0$ it is 3200--5600 at 10~MeV and 370--670 at 20~MeV for $Y_L=0.3$, 390--1150 and 86--160 for $Y_L=0.1$, and up to $4\times10^4$ at 5~MeV. The temperature dependence comes almost entirely from the nucleonic mean free path, which grows rapidly as the temperature, and with it the energy of the diffusing antineutrinos, decreases. The mean free path with $\Delta$s, which is set by the three-body opacity, depends little on temperature: at 2--$3n_0$ it is 0.7--1.3~cm at $Y_L=0.3$ and 2.6--16~cm at $Y_L=0.1$. Without the three-body channel the ratio would be 1.0--2.4, from neutral-current scattering on the $\Delta^-$ and the weaker degeneracy of the $\Delta^-$ compared to the neutrons they replace. A width of the final $\Delta$ (open diamonds and plus signs) lengthens the mean free path with $\Delta$s: $W=T^2/T_W$ by 5--10\% at 10~MeV and 11--37\% at 20~MeV, and the vacuum width of 117~MeV by 18--61\%. The ratio then remains above 60.

Merger remnants reach temperatures above 20~MeV, the highest ones near saturation density and up to a few tens of MeV in the compressed core~\cite{Perego2019}, and 30--50~MeV in the most strongly shocked regions~\cite{Sedrakian:2026lfd}. We extended the quasiparticle calculation to $T=30$ and 50~MeV at $3n_0$; there the collisional width $T^2/T_W$ would exceed the Fermi energies, and we do not apply it. For neutrinos the ratio is 1.9--2.9 at $Y_L=0.3$ and 2.6--5.9 at $Y_L=0.1$, similar to its values at 20~MeV. For antineutrinos the three-body channel remains open, with threshold margins of 70--270~MeV, and the mean free path with $\Delta$s stays at 0.9--3.6~cm. The nucleonic mean free path falls to 1.2--1.3~m at 30~MeV and to 0.25~m at 50~MeV, so the ratio decreases to 33--150 at 30~MeV and to 8--20 at 50~MeV. The antineutrino effect thus weakens with temperature but remains an order of magnitude at 50~MeV.

\subsection{Scaling with the $\Delta$ fraction and with the $\Delta^-$ threshold}
\label{sec:scaling}
\begin{figure*}[tbp]
\includegraphics[width=\textwidth]{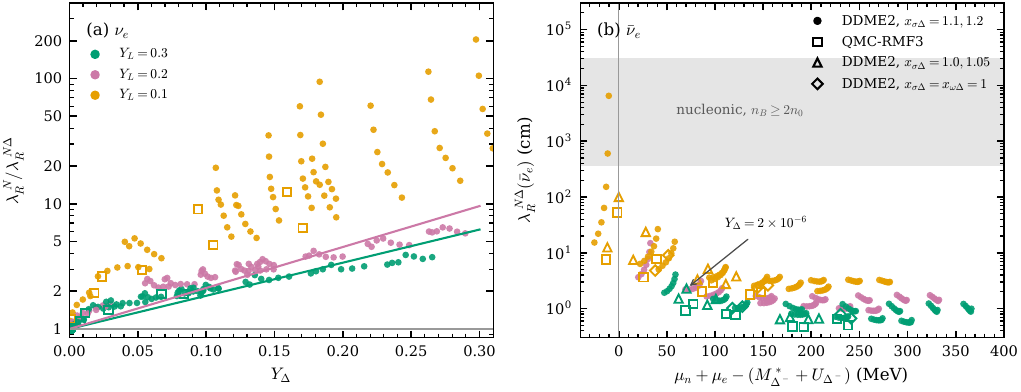}
\caption{(a) Ratio $\lambda_R^N/\lambda_R^{N\Delta}$ for electron neutrinos as a function of the total $\Delta$ fraction $Y_\Delta$. (b) Antineutrino diffusion mean free path with $\Delta$s, including three-body absorption, as a function of the threshold margin $\mu_n+\mu_e-(M^*_{\Delta^-}+U_{\Delta^-})$ of Eq.~\eqref{eq:thr}. Dots: DDME2 ($x_{\sigma\Delta}=1.1$ and $1.2$; $n_B=1.5$--$4n_0$; $T=5$--$20$~MeV), colored by lepton fraction. Open squares: QMC-RMF3 ($n_B=2$ and $3n_0$; $T=10$ and $20$~MeV), not used in the fits. Open triangles in (b): DDME2 with $x_{\sigma\Delta}=1.0$ and 1.05 ($x_{\omega\Delta}=1.1$); open diamonds: $x_{\sigma\Delta}=x_{\omega\Delta}=1$, both at $n_B=2$ and $3n_0$ and $T=10$ and 20~MeV. Shaded band in (b): range of the nucleonic mean free path at $n_B\geq2n_0$. Each streak of dots is a temperature sequence (5--20~MeV) at fixed model, density and $Y_L$. The arrow marks the state with $Y_\Delta=2\times10^{-6}$ discussed in the text. Lines in (a): fits of Table~\ref{tab:fits}. Quasiparticle approximation. Except at $Y_L=0.1$, the neutrino enhancement is set by the $\Delta$ fraction; the antineutrino mean free path is set by the threshold margin and is \nubarLamDMin--\nubarLamDMax~cm once the margin exceeds about 50~MeV.}
\label{fig:scaling}
\end{figure*}

The ratio depends on density, temperature, lepton fraction and the $\Delta$ couplings, and we would like to know whether a single variable captures most of this dependence. Figure~\ref{fig:scaling}(a) shows the neutrino ratio on a grid of densities ($1.5$--$4n_0$), temperatures ($5$--$20$~MeV) and lepton fractions (0.1, 0.2, 0.3) for DDME2 with both coupling sets, plotted against the total $\Delta$ fraction $Y_\Delta$. Points for QMC-RMF3 are added, which were not used in the fits.

For electron neutrinos at $Y_L=0.2$ and 0.3 the ratio grows exponentially with $Y_\Delta$, with a slope that increases as the lepton fraction decreases [Fig.~\ref{fig:scaling}(a)]. The exponential form describes the data to 17--21\% rms; it slightly overestimates the ratio at the largest $\Delta$ fractions. At $Y_L=0.1$ the points scatter by an order of magnitude at fixed $Y_\Delta$, because there the ratio is controlled by the opening of nucleonic absorption, which depends on how far the proton fraction is pushed above the direct Urca threshold and on the temperature, not on the $\Delta$ fraction alone.

For antineutrinos the $\Delta$ fraction is not the controlling variable, because the three-body channel does not require $\Delta$s in the matter. Figure~\ref{fig:scaling}(b) shows the antineutrino mean free path with $\Delta$s against the threshold margin of Eq.~\eqref{eq:thr}. Once the margin exceeds about 50~MeV, $\lambda_R$ lies between \nubarLamHiYLmin\ and \nubarLamHiYLmax~cm at $Y_L=0.2$--0.3 and between \nubarLamLoYLmin\ and \nubarLamLoYLmax~cm at $Y_L=0.1$, for both equations of state and all coupling sets (the lowest values are those of QMC-RMF3; for DDME2 0.56--4 and 2.1--16~cm), compared to 4--300~m in nucleonic DDME2 matter at $n_B\geq2n_0$ (shaded band). It is longer at low lepton fraction, where the electron density is lower. The three-body channel supplies more than 94\% of the opacity at all these states. Below threshold $\lambda_R$ rises steeply, but the channel still operates through the thermal tails: at margins of $-10$ to $-25$~MeV ($x_{\sigma\Delta}=1.1$, $1.5n_0$, $Y_L=0.1$) the mean free path is 15~cm at 20~MeV, compared to 410~cm in nucleonic matter, and 65~m at 5~MeV, compared to 320~m. The open triangles, for $x_{\sigma\Delta}=1.0$ and 1.05, include states where the $\Delta$ fraction is as small as $2\times10^{-6}$ ($x_{\sigma\Delta}=1.0$, $2n_0$, $Y_L=0.3$, $T=10$~MeV); the margin there is 70~MeV and the mean free path is shorter than in nucleonic matter by a factor of 1600. The open diamonds are for $x_{\sigma\Delta}=x_{\omega\Delta}=1$, for which the $\Delta$ potential in symmetric matter equals the nucleon one, at the upper edge of the empirical range quoted in Sec.~\ref{sec:matter}. They fall on the same band: $\lambda_R=0.7$--1.1~cm at $Y_L=0.3$ and 2.4--9~cm at $Y_L=0.1$ (2--$3n_0$, 10--20~MeV). The result therefore does not rest on the strongly attractive coupling $x_{\sigma\Delta}=1.2$. A fit in $Y_\Delta$ is not meaningful for antineutrinos. What determines their opacity is the margin, and with it the $\Delta^-$ potential in dense matter.

\begin{table}[tbp]
\caption{Fits of $\lambda_R^N/\lambda_R^{N\Delta}$ for electron neutrinos as a function of the total $\Delta$ fraction, to the DDME2 grid (quasiparticles, $x_{\sigma\Delta}=1.1$ and $1.2$, $1.5$--$4n_0$, $5$--$20$~MeV). The rms deviation is in the natural logarithm of the ratio; the last column gives it for the QMC-RMF3 points, which were not fitted.}
\label{tab:fits}
\begin{ruledtabular}
\begin{tabular}{llcc}
Species, $Y_L$ & Fit & rms & QMC-RMF3 \\
\hline
$\nu_e$, $0.3$ & $\exp(a\,Y_\Delta)$, $a=6.09$ & 0.17 & 0.14 \\
$\nu_e$, $0.2$ & $\exp(a\,Y_\Delta)$, $a=7.53$ & 0.21 & -- \\
$\nu_e$, $0.1$ & no single-variable fit & -- & -- \\
\end{tabular}
\end{ruledtabular}
\end{table}

\subsection{Antineutrino share of lepton-number and energy diffusion}
\label{sec:share}
In the diffusion approximation the part of the lepton-number flux driven by the gradient of the neutrino degeneracy has as coefficient the sum of the neutrino and antineutrino coefficients $D_2$~\cite{Pons:1998mm}; the remaining part, driven by the temperature gradient, involves $D_3$. For each species $D_2$ is the product of $\lambda_R$ and the normalization integral $\int dE\,E^2f_\nu(1-f_\nu)$ of Eq.~\eqref{eq:lamR}. When $\mu_{\nu_e}\gg T$ the antineutrino integral is exponentially small, and antineutrinos carry little lepton number however long their mean free path. At $Y_L=0.3$ and $n_B\geq2n_0$ the antineutrino share of the total $D_2$ is below $4\times10^{-4}$ even without the three-body channel, and at $Y_L=0.2$ about $10^{-2}$. There the reduction of $\lambda_R(\bar\nu_e)$ changes lepton-number transport by less than 2\%.

\begin{figure*}[tbp]
\includegraphics[width=\textwidth]{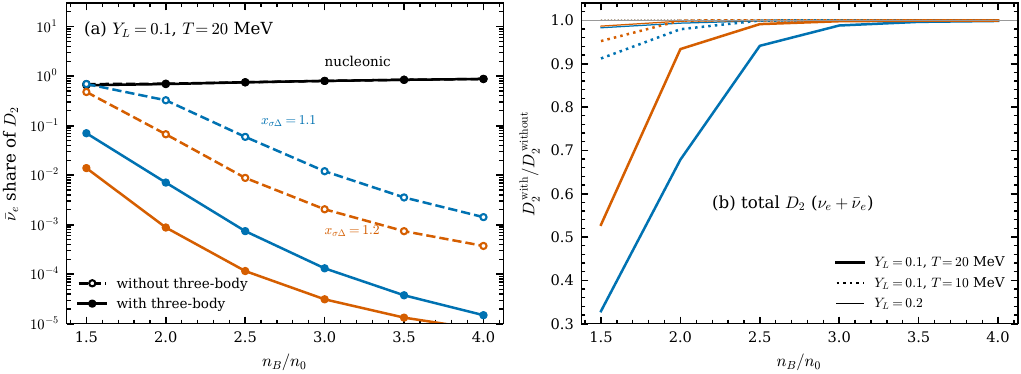}
\caption{Lepton-number diffusion. (a) Antineutrino share of the total lepton-number diffusion coefficient $D_2=D_2^{\nu_e}+D_2^{\bar\nu_e}$ at $Y_L=0.1$ and $T=20$~MeV, as a function of baryon density, without (dashed, open circles) and with (solid, filled circles) three-body absorption, for nucleonic DDME2 (black) and with $\Delta$s for $x_{\sigma\Delta}=1.1$ (blue) and 1.2 (orange). (b) Ratio of the total $D_2$ with and without the three-body channel, for $Y_L=0.1$ at $T=20$~MeV (solid) and 10~MeV (dotted), and for $Y_L=0.2$ at $T=20$~MeV (thin). Quasiparticle approximation. The channel matters for lepton-number transport only where antineutrinos are abundant, at low lepton fraction and low density.}
\label{fig:d2}
\end{figure*}

At $Y_L=0.1$ the neutrino chemical potential is comparable to a few $T$ near the $\Delta$ threshold (Fig.~\ref{fig:comp}), and the antineutrinos matter. Without the three-body channel they would carry up to 33\% of $D_2$ at $n_B\geq2n_0$ ($x_{\sigma\Delta}=1.1$, $2n_0$, $T=20$~MeV, $\mu_{\nu_e}=51$~MeV) and up to 69\% at $1.5n_0$; with it their share falls below 1\% and to 7\%, respectively. The total lepton-number diffusion coefficient then decreases by up to \DtwoRedHi\% at $n_B\geq2n_0$ and by a factor of up to \DtwoRedLow\ at $1.5n_0$, $T=20$~MeV (Fig.~\ref{fig:d2}). At $Y_L=0.2$ it is below \DtwoRedYLtwo\%. In nucleonic matter at $Y_L=0.1$, where $\mu_{\nu_e}$ nearly vanishes at high density, the antineutrino share reaches 88\%; if the $\Delta$ resonance is accessible in such matter, this is where the three-body channel would change transport most. The part of the energy flux driven by the temperature gradient has the coefficient $D_4=D_4^{\nu_e}+D_4^{\bar\nu_e}+4D_4^{\nu_x}$ of Ref.~\cite{Pons:1998mm}, where $D_n$ weights the energy-dependent diffusion coefficient with $x^nf_\nu(1-f_\nu)$, $x=E/T$, and $\nu_x$ stands for the four heavy-lepton neutrino species (the energy flux also contains a term driven by the degeneracy gradient, with coefficient $D_3$). We computed $D_4$ at $2$ and $3n_0$, $T=10$ and 20~MeV, with the heavy-lepton neutrinos at zero chemical potential and opacities from neutral-current scattering. At $Y_L=0.3$ electron neutrinos carry 79--99\% of $D_4$ and heavy-lepton neutrinos the rest; at $Y_L=0.1$ and 20~MeV heavy-lepton neutrinos carry 34--85\%. Electron antineutrinos carry at most 2\% of $D_4$ in $\Delta$-admixed matter even without the three-body channel, so the channel changes $D_4$ by less than 2\%. The $\Delta$s themselves change $D_4$ by factors of 0.8--1.0 at $Y_L=0.3$; at $Y_L=0.1$, $3n_0$ and 10~MeV they raise it by factors of up to five, because the neutrino chemical potential they induce increases the number of diffusing neutrinos more than the opacity grows.

\subsection{In-medium width of the $\Delta^-$ and stability against pion condensation}
\label{sec:pion}

\begin{figure*}[tbp]
\includegraphics[width=\textwidth]{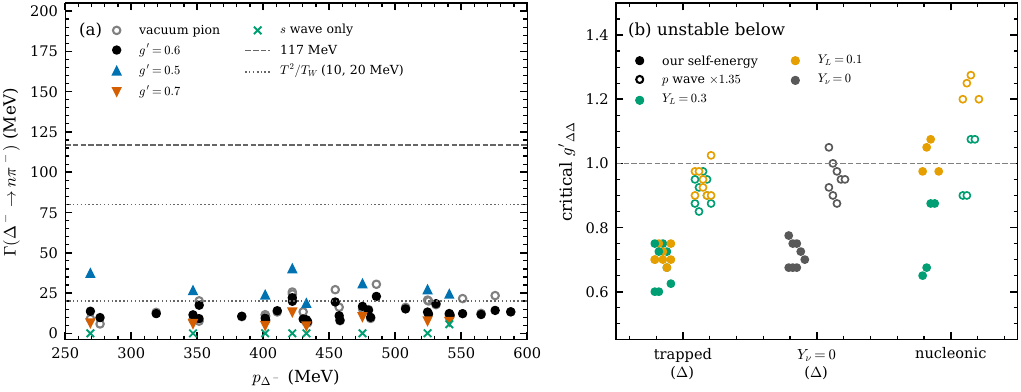}
\caption{(a) $N\pi$ width of the $\Delta^-$ produced in $\bar\nu_e+e^-+n\to\Delta^-$, Eq.~\eqref{eq:Gmed}, as a function of its momentum, at sixteen states (DDME2, $x_{\sigma\Delta}=1.1$ and 1.2, $Y_L=0.1$ and 0.3, 2 and $3n_0$, $T=10$ and 20~MeV; antineutrino energies $T$ and $3T$): vacuum pion propagator (open circles), in-medium propagator with $g'=0.6$ (filled circles), and, at $T=20$~MeV, $g'=0.5$ and 0.7 and the $s$-wave self-energy alone. Horizontal lines: the vacuum Breit--Wigner width and the collisional width $T^2/T_W$ at 10 and 20~MeV. (b) Critical $g'_{\Delta\Delta}$ for $p$-wave $\pi^-$ condensation, with $g'_{NN}=0.59$ and $g'_{N\Delta}=0.16+0.06\,g'_{\Delta\Delta}$~\cite{SuzukiSakaiTatsumi1999}, for trapped $\Delta$-admixed matter (16 states), neutrino-free $\Delta$-admixed matter (8 states) and nucleonic matter (8 states) at the same density and temperature; the matter is unstable below the plotted value. Filled: our self-energy; open: squared $p$-wave couplings multiplied by 1.35, which reproduces the critical density of Ref.~\cite{SuzukiSakaiTatsumi1999}. Colors: lepton fraction (gray: neutrino-free). Dashed line: $g'_{\Delta\Delta}=1$, the upper value assumed in Ref.~\cite{SuzukiSakaiTatsumi1999}. The values are upper bounds on a grid of step 0.025.}
\label{fig:pion}
\end{figure*}

The repulsive $s$-wave term raises the $\pi^-$ mass to 156--194~MeV at $k=0$; the $p$-wave attraction lowers the pion branch at $k\approx200$--300~MeV, but for $g'\geq0.5$ it stays at least 15~MeV above $\mu_\pi$ at the states we examined (2--$3n_0$, $T=10$~MeV, both models and lepton fractions), so there is no $\pi^-$ condensation in this model. With $g'=0.6$ the width of the produced $\Delta^-$ is \GamRefMin--\GamRefMax~MeV [Fig.~\ref{fig:pion}(a)], between \GamRatioMin\ and \GamRatioMax\ times the vacuum-pion value, and $g'=0.5$ and 0.7 give 19--40~MeV and 4.5--13~MeV at $T=20$~MeV. The $s$-wave term alone would reduce the width below 6~MeV by closing the phase space; the $p$-wave attraction reopens it. The width assigned to the $\Delta$ in the $\Delta$-hole term (20--117~MeV) has no effect, because in the relevant kinematics the $\Delta$-hole states lie below the $N\pi$ threshold. The model extrapolates chiral input to 2--$3n_0$, above about $1.5n_0$, where the convergence of chiral perturbation theory for the pion self-energy becomes an issue in symmetric matter and the uncertainty from poorly constrained two-pion--two-nucleon couplings grows rapidly~\cite{ForeKaiserReddyWarrington2024}, and uses a single Landau--Migdal parameter, a simplification of the separate $NN$, $N\Delta$ and $\Delta\Delta$ parameters of Ref.~\cite{HelgessonRandrup1995}. We take the spread over $g'$ as its uncertainty. This spread does not cover all choices. With the separate parameters $g'_{NN}=0.9$, $g'_{N\Delta}=0.38$ and $g'_{\Delta\Delta}=0.35$ of Ref.~\cite{HelgessonRandrup1995}, used in the two-channel form of the RPA, the weaker screening of the $\Delta$-hole polarization brings the pion branch down to or below $\mu_\pi$ at seven of our eight states at $T=20$~MeV; the model then predicts $p$-wave $\pi^-$ condensation, which our quasiparticle treatment does not describe. At the remaining state ($x_{\sigma\Delta}=1.1$, $Y_L=0.3$, $2n_0$) the width is 111~MeV, close to the vacuum value. The in-medium width of the produced $\Delta^-$, and the stability of hot $\Delta$-admixed matter against pion condensation, therefore depend on the poorly known short-range $N\Delta$ and $\Delta\Delta$ interactions. To locate the instability we scanned, at all sixteen states with $x_{\sigma\Delta}=1.1$ and 1.2, $Y_L=0.1$ and 0.3, 2 and $3n_0$, and $T=10$ and 20~MeV, the minimum over momentum of the $\pi^-$ branch relative to $\mu_\pi$, with $g'_{NN}=0.6$ or 0.9 and $g'_{N\Delta}=g'_{\Delta\Delta}\equiv g'_\Delta$. The branch reaches $\mu_\pi$, the onset of $p$-wave $\pi^-$ condensation, when $g'_\Delta$ falls below 0.33--0.45; for $g'_\Delta\geq0.5$ it stays at least 17~MeV above $\mu_\pi$. The Gamow--Teller quenching measured on $^{90}$Zr fixes $g'_{NN}\approx0.59$ and $g'_{N\Delta}=0.16+0.06\,g'_{\Delta\Delta}$ but leaves $g'_{\Delta\Delta}$ open; Ref.~\cite{SuzukiSakaiTatsumi1999} assumed $g'_{\Delta\Delta}<1$. As a test of our self-energy we repeated their calculation of $\pi^0$ condensation in symmetric nuclear matter (static limit, $m^*=0.8M$, cutoff 1~GeV; densities in units of their $\rho_0=0.17$~fm$^{-3}$). We obtained a critical density of $2.55\rho_0$ for $g'_{\Delta\Delta}=1$ ($2.26\rho_0$ without the recoil of the $\Delta$), where they found $1.6\rho_0$; with a universal $g'$ we found $3.1\rho_0$ for $g'=0.5$, $10\rho_0$ for $g'=0.6$ and no condensation up to $12\rho_0$ for $g'=0.7$ and 0.8, as expected: including the form factor in their high-density condition, we estimate that it cannot be met for $g'>0.665$. Our $p$-wave attraction is therefore weaker than theirs. Multiplying both squared $p$-wave couplings by 1.35 reproduces their $1.6\rho_0$, and we used this as a second variant. With the Gamow--Teller values we then located the critical $g'_{\Delta\Delta}$ from the condition $\max_k\mathrm{Re}\,D^{-1}(\mu_\pi,k)=0$, the first two threshold conditions of Ref.~\cite{SuzukiSakaiTatsumi1999} with $\mu_\pi$ fixed by $\beta$ equilibrium, which agrees with the branch criterion and is monotonic in $g'_{\Delta\Delta}$ at every state. On a grid of step 0.025 in $g'_{\Delta\Delta}$, the trapped matter is stable for $g'_{\Delta\Delta}$ above \gcritTr [Fig.~\ref{fig:pion}(b)] with our self-energy and above \gcritTrSc in the rescaled variant; at $g'_{\Delta\Delta}=1$ the $\pi^-$ branch stays 42--83~MeV above $\mu_\pi$ with our self-energy. Neutrino trapping has little effect: in neutrino-free matter at the same density and temperature the critical values are \gcritNf and \gcritNfSc. The instability is not caused by the $\Delta$ population. Nucleonic matter at the same density, temperature and lepton fraction, with virtual $\Delta$-hole states, requires $g'_{\Delta\Delta}$ above \gcritNuc, or \gcritNucSc in the rescaled variant, because at $Y_L=0.1$ its $\mu_\pi$ is \mupiNucLoYL~MeV, whereas the $\Delta^-$, by replacing electrons, lowers it to \mupiDLoYL~MeV. Whether dense matter condenses with the Gamow--Teller values of $g'_{NN}$ and $g'_{N\Delta}$ therefore depends on $g'_{\Delta\Delta}$ and on the normalization of the $p$-wave attraction, and it applies to our nucleonic reference states at least as much as to the $\Delta$-admixed ones. Like nucleonic opacity calculations, our results apply to the uncondensed phase.

\subsection{Sensitivity to the $\Delta$ weak inputs and to correlations}
\label{sec:sens}
Table~\ref{tab:sens} lists the ratio at $3n_0$ and $T=20$~MeV for DDME2 with $x_{\sigma\Delta}=1.2$ under variations of the inputs. The $\Delta$ axial coupling $g_{A\Delta}$ is the least known input~\cite{JiangTiburzi2008,Jun2020}; published values are given in different normalizations, and we vary it between zero and $g_A$. Setting it to zero removes the axial part of the $\Delta\to\Delta'$ and $\Delta$ neutral-current reactions; the neutrino enhancement is reduced to 2.1 and 7.6 but remains, because the vector couplings are fixed by current conservation and the $\Delta N$ transitions do not involve $g_{A\Delta}$. For antineutrinos $g_{A\Delta}$ changes the ratio by at most 4\%. The $N\Delta$ vertex enters mainly through the antineutrino ratio, which is set by the three-body channel. Reducing $C_5^A(0)$ from 1.2 to the fit value 0.867 of Ref.~\cite{HNV2007} lowers it by a factor of 1.67, close to $(1.2/0.867)^2=1.92$ reduced by the vector contribution, and lowers the neutrino ratio by 8--9\%. The fit of Ref.~\cite{HNV2007} pairs $C_5^A(0)=0.867$ with the axial mass $M_A=0.985$~GeV; with this pair the antineutrino ratio at $Y_L=0.3$ is 239 instead of 236, and the axial mass alone changes the reference value by 2\%. The $C_5^A$-only vertex of Ref.~\cite{Kumar2026P1} gives an antineutrino ratio lower by 28--33\%, and dropping only the vector form factors lowers it by 22--25\%: the vector $N\Delta$ transition, fixed by electroproduction, raises the three-body rate by about a third. The vacuum width of the final $\Delta$ lowers the antineutrino ratio by 15--23\%.

For antineutrinos the many-body treatments act mostly on the nucleonic reference, and we separate the two effects in Table~\ref{tab:sens}. With the nucleonic mean free path held at its quasiparticle value, the treatments change the antineutrino ratio by $-11\%$ to $-1\%$ at $Y_L=0.3$ and by $-18\%$ to $-1\%$ at $Y_L=0.1$: the mean free path with $\Delta$s changes by at most 21\%. The nucleonic mean free path, set by neutral-current scattering, grows by a factor of 1.6 in the NWA, which is not appropriate for neutral currents~\cite{AHZ2024nwa}, shrinks by a factor of about three with the energy broadening of Eq.~\eqref{eq:SG}, and grows by 13--20\% with the RPA. When each treatment is applied to both kinds of matter, the ratio therefore ranges from 38 to 570 at this state (numbers in parentheses); this spread is a property of the nucleonic scattering response, not of the $\Delta$s.

Long-wavelength correlations reduce the vector response by factors $R_V\approx0.4$--0.7, larger in nucleonic than in $\Delta$-admixed matter. They lengthen the mean free paths by 1--8\% for neutrinos, whose opacity is dominated by absorption, and change the neutrino ratio by at most a few percent. For antineutrinos they act only on the scattering opacity, which with $\Delta$s supplies, together with two-body absorption, 2--4\% of the total at the states of Table~\ref{tab:sens}. Collisional broadening was discussed above.

\begin{table}[tbp]
\caption{Sensitivity of $\lambda_R^N/\lambda_R^{N\Delta}$ at $n_B=3n_0$, $T=20$~MeV, DDME2 with $x_{\sigma\Delta}=1.2$. The reference row uses the inputs of Table~\ref{tab:inputs} and quasiparticles. Full $N\Delta$ vertex. NWA: nucleon width approximation for all two-body reactions; NWA+NC: width approximation for absorption and energy broadening for scattering; both with the final-$\Delta$ width $T^2/T_W$ in the three-body channel. RPA: long-wavelength vector correlations. Antineutrino ratios include three-body absorption. For the antineutrino entries of the NWA, NWA+NC and RPA rows the nucleonic mean free path is kept at its quasiparticle value, so that the entry measures the effect of the treatment on $\Delta$-admixed matter; the ratio with the treatment applied to both kinds of matter is given in parentheses.}
\label{tab:sens}
\begin{ruledtabular}
\begin{tabular}{lcccc}
 & \multicolumn{2}{c}{$Y_L=0.3$} & \multicolumn{2}{c}{$Y_L=0.1$} \\
Variation & $\nu_e$ & $\bar\nu_e$ & $\nu_e$ & $\bar\nu_e$ \\
\hline
Reference & 2.63 & 393 & 10.6 & 128 \\
$g_{A\Delta}=0$ & 2.06 & 389 & 7.61 & 126 \\
$g_{A\Delta}=g_A$ & 3.36 & 403 & 14.7 & 133 \\
$C_5^A(0)=0.867$ & 2.43 & 236 & 9.67 & 76.5 \\
$C_5^A$-only vertex~\cite{Kumar2026P1} & 2.47 & 263 & 9.77 & 91.5 \\
No vector $N\Delta$ form factors & 2.29 & 294 & 9.33 & 99.4 \\
NWA & 5.13 & 351 (574) & 9.73 & 105 (173) \\
NWA+NC & 3.99 & 358 (123) & 9.70 & 111 (38) \\
RPA & 2.62 & 391 (441) & 10.7 & 126 (152) \\
Final-$\Delta$ width 117~MeV & -- & 334 & -- & 99.0 \\
\end{tabular}
\end{ruledtabular}
\end{table}

\section{Discussion and conclusions}
\label{sec:concl}
We have calculated the diffusion mean free paths of electron neutrinos and antineutrinos in hot, neutrino-trapped matter containing $\Delta(1232)$ isobars, with all charged-current and neutral-current channels of the $\Delta$ quartet, three-body antineutrino absorption, relativistic matrix elements and mean-field kinematics. Our main results are as follows.

First, the three-body reaction $\bar\nu_e+e^-+n\to\Delta^-$ dominates the antineutrino opacity wherever the threshold condition $\mu_n+\mu_e>M^*_{\Delta^-}+U_{\Delta^-}$ holds. At 2--$4n_0$ it shortens the antineutrino diffusion mean free path from \nubarLamNucMin--\nubarLamNucMax~m in nucleonic matter to \nubarLamDMin--\nubarLamDMax~cm where the reaction is open, by factors of \nubarRatioMin\ at 20~MeV to $\nubarRatioMaxMant\times10^{\nubarRatioMaxExp}$ at 5~MeV. The reaction does not require a $\Delta$ population: it reduces the mean free path by a factor of \smallYDratio\ at a state with $Y_\Delta=2\times10^{-6}$. It does not rest on a strongly attractive $\Delta$ potential either: for a $\Delta$ potential equal to the nucleon one the mean free path is \xoneLamMin--\xoneLamMax~cm.

Second, the $\Delta$s shorten the neutrino diffusion mean free path by factors of up to \nuRatioYLhiMax\ at $Y_L=0.2$--0.3 and $2$--$4n_0$, increasing with the $\Delta$ fraction, and by factors of \nuRatioYLoneMin--\nuRatioYLoneMax\ at $Y_L=0.1$. At $Y_L\geq0.2$ the cause is absorption on the $\Delta^-$; at $Y_L=0.1$ the $\Delta^-$ in addition raises the proton fraction above the direct Urca threshold and raises the neutrino chemical potential, which opens nucleonic absorption. At 2--$3n_0$ the two-body $\Delta$ reactions alone would shorten the antineutrino mean free path by factors of up to \nubarTwoBodyMax, through neutral-current scattering on the $\Delta^-$, whose large weak vector charge is fixed by current conservation.

Third, the static limit for spin-$\tfrac32$ targets shows that, for a given axial coupling in the vertex, the $\Delta$ spin reduces the Gamow--Teller strength per particle from 3 to $5/3$ in units of the axial coupling squared. The importance of the $\Delta^-$ for scattering comes from its vector charge, not from its spin.

Fourth, for neutrinos at $Y_L\geq0.2$ the ratio of the mean free paths is a function of the total $\Delta$ fraction that depends only weakly on the equation of state, and a simulation can correct nucleonic opacities with the fits of Table~\ref{tab:fits}. At low lepton fraction the correction must be computed from the opacities directly. Tables of $\lambda_R$ for both species on the DDME2 grid, with and without the three-body channel and with the threshold margin, are provided as Supplemental Material. For antineutrinos the controlling variable is the threshold margin, which an equation-of-state table can supply from the neutron and electron chemical potentials and the $\Delta^-$ self-energies.

Fifth, the neutrino enhancement remains when the poorly known $\Delta$ axial coupling is set to zero, changes by about ten percent over the range of $C_5^A$, and changes by a few percent with long-wavelength correlations; collisional broadening reduces the near-threshold enhancement at low lepton fraction by up to a factor of two but leaves it at 6--15 at $3n_0$. The antineutrino result scales approximately with $C_5^{A\,2}$; the vector $N\Delta$ form factors, which Ref.~\cite{Kumar2026P1} neglected, raise it by about a third, and a width of the final $\Delta$ up to the vacuum one lowers it by 15--23\%. The in-medium $\Delta^-\to n\pi^-$ width of the produced $\Delta^-$, \GamMedMin--\GamMedMax~MeV with the in-medium pion over the range of $g'$, lies inside this range. In all variations the antineutrino mean free path remains shorter than the nucleonic one by more than an order of magnitude.

In a diffusion description the lepton-number diffusion time scales inversely with $\lambda_R$. If $\Delta$s are present in the core of a proto-neutron star, our results imply that lepton number carried by neutrinos diffuses more slowly than in nucleonic matter at the same temperature, in the regions where the $\Delta$s are abundant. Antineutrinos with mean free paths of centimeters are in local equilibrium with the matter, and their diffusive contribution to transport in the hot, dense regions is suppressed by the same large factors. This matters only where antineutrinos are abundant, that is where $\mu_{\nu_e}$ is at most a few $T$: at $Y_L=0.2$--0.3 they carry less than 1\% of the lepton-number diffusion even without the three-body channel, whereas at $Y_L=0.1$ the channel lowers the total lepton-number diffusion coefficient by up to \DtwoRedHi\% at $n_B\geq2n_0$ and by a factor of up to \DtwoRedLow\ at $1.5n_0$ (Sec.~\ref{sec:share}). Energy transport, carried by electron and heavy-lepton neutrinos, changes by less than 2\%. Late deleptonization of proto-neutron stars and merger remnants, where the lepton fraction is low, are therefore the conditions in which the channel should be tested in transport calculations. How much this changes the deleptonization time is not settled by the opacities alone. In evolution calculations with hyperons, which raise the cross sections by factors of two to five at fixed temperature, the evolution timescale became only slightly longer~\cite{Pons:1998mm}. The same work showed that the additional baryons raise the specific heat, so that at fixed entropy the temperature is lower and the trend of the opacities can reverse~\cite{Pons:1998mm}; a similar competition is to be expected for $\Delta$s. In merger remnants, where neutrinos in the hot, dense regions are close to thermal equilibrium with the matter~\cite{Alford:2026kwd}, the shorter mean free paths would delay the release of the trapped lepton number and energy. Quantifying these effects requires transport simulations with $\Delta$-admixed opacities, including the three-body channel.

Our calculation has several limitations. We have not included correlations in the spin channel, which are not constrained by the mean-field models and are a major uncertainty already for nucleonic matter at and below saturation density~\cite{Lin:2022lug} and change the charged-current rates qualitatively at subsaturation densities~\cite{Shin:2023sei}, Coulomb correlations, or heavy-lepton neutrino opacities other than neutral-current scattering on baryons, and we treat temperatures above 20~MeV, which are reached in merger remnants mainly near and somewhat above saturation density~\cite{Perego2019}, only in the quasiparticle approximation at $3n_0$. The three-body channel is computed with a quasiparticle $\Delta$ in the final state. The $\Delta^-$ is produced with momenta of 250--590~MeV, far above its Fermi surface, where the accuracy of the mean-field dispersion relation is not established; our in-medium $N\pi$ width uses a pion self-energy extrapolated to these densities, depends on $g'$ by a factor of three to six, and with the Gamow--Teller values of $g'_{NN}$ and $g'_{N\Delta}$ the matter, nucleonic or $\Delta$-admixed, is stable against $p$-wave $\pi^-$ condensation only for $g'_{\Delta\Delta}\gtrsim0.6$--1, and the vacuum width lengthens the antineutrino mean free path by up to 60\%. A complete treatment would compute the in-medium $N\pi$ continuum, including nonresonant pion production, together with the resonance. The threshold condition depends on the $\Delta^-$ potential at high density, which is poorly constrained. Our nucleonic reference omits the $\Delta$ resonance altogether; in nucleonic matter with the resonance at a comparable position the channel would be open as well, so the antineutrino ratios we quote compare matter with and without the resonance, not with and without a $\Delta$ population. The $N\Delta$ form factors are taken from vacuum weak pion production~\cite{HNV2007}; their modification in dense matter is not known, and $C_5^A(0)$ itself differs by about 30\% between the off-diagonal Goldberger--Treiman value, 1.15 (1.2 in the parametrization we use), and the fit to the Argonne bubble-chamber data, 0.867 with a smaller axial mass~\cite{HNV2007}. The collisional width is a parametrization, and at 20~MeV it is large enough that the broadened results are only indicative. Finally, the abundance and the potential of $\Delta$s in dense matter are themselves uncertain; the neutrino opacity depends mainly on the abundance and the antineutrino opacity on the potential, so constraints on either translate directly into constraints on the opacities.

\begin{acknowledgments}
The author thanks the National Institute of Technology Rourkela for support. 
\end{acknowledgments}

\appendix
\section{Phase-space reduction}
\label{app:ps}
We integrate Eq.~\eqref{eq:lambda} over $\mathbf p_f$ with the spatial part of the delta function, which sets $\mathbf p_f=\mathbf k+\mathbf p_i-\mathbf q$. The remaining delta function is $\delta(h(q))$ with $h$ of Eq.~\eqref{eq:h}. Writing $d^3q=q^2\,dq\,d\Omega_q$,
\begin{equation}
\delta\big(h(q)\big)=\sum_{\text{roots }q_r}\frac{\delta(q-q_r)}{|h'(q_r)|} .
\label{eq:deltah}
\end{equation}
The derivative of $h$ is Eq.~\eqref{eq:hp}. For the initial baryon we use $E_i^*\,dE_i^*=p_i\,dp_i$ and $dE_i=dE_i^*$ at fixed density, so that
\begin{equation}
\frac{d^3p_i}{2E_i^*}=\frac{p_i}{2}\,dE_i\,d\cos\theta_i\,d\phi_i .
\label{eq:dpi}
\end{equation}
With $\mathbf k$ along $z$ the integrand does not depend on $\phi_i$ once the lepton direction is integrated over the full sphere, so $\int d\phi_i=2\pi$. Collecting the factors $(2\pi)^4/(2\pi)^9$ and $1/8$ from the three factors $1/(2E^*)$ gives Eq.~\eqref{eq:lambda4}.

Both $E_f^*(q)=[|\mathbf Q-q\hat{\mathbf q}|^2+m_f^{*2}]^{1/2}$, with $\mathbf Q=\mathbf k+\mathbf p_i$, and $E_q=(q^2+m_l^2)^{1/2}$ are convex functions of $q$: the first is a nondecreasing convex function of the norm of an affine function of $q$, and the second is a hyperbola. Hence $h$ is convex and has at most two roots on $q\geq0$. We find the minimum $q_{\rm min}$ of $h$ by bisection on $h'$, or set $q_{\rm min}=0$ if $h'(0)\geq0$. If $h(q_{\rm min})\geq0$ there is no root. Otherwise there is one root above $q_{\rm min}$, and a second one in $(0,q_{\rm min})$ if $h(0)>0$; both are found by bisection to machine precision.

The integration variables are $E_i$, $\cos\theta_i$ and the two angles of $\hat{\mathbf q}$. The initial energy is restricted to $\max(U_i+m_i^*,\,\mu_i-40T)\leq E_i\leq\max(\mu_i,\,U_i+m_i^*)+40T$, outside which $f_i$ or the final-state blocking makes the integrand negligible. We map the four variables to the unit hypercube and sample it with scrambled Sobol sequences of $2^{13}$ to $2^{16}$ points (Appendix~\ref{app:conv}). For neutral-current scattering the outgoing lepton is a neutrino, $m_l=0$, and the baryon is unchanged; for the transport opacity the integrand carries the additional factor $1-\cos\theta_{kq}$.

\textit{Three-body absorption.} In Eq.~\eqref{eq:lamID} we integrate over $\mathbf p_L$ with the spatial delta function, which sets $\mathbf p_L=\mathbf Q+\mathbf p_H$. We write $d^3p_H=p_H^2\,dp_H\,d\cos\alpha\,d\phi$ with the polar axis along $\mathbf Q$. Since
\begin{equation}
E_L^{*2}=Q^2+p_H^2+2Qp_H\cos\alpha+m_L^{*2},\qquad \frac{\partial E_L^*}{\partial\cos\alpha}=\frac{Qp_H}{E_L^*} ,
\label{eq:dEL}
\end{equation}
the energy delta function $\delta(E+E_e+E_H-U_L-E_L^*)$ fixes $\cos\alpha$ to Eq.~\eqref{eq:cosa} and contributes the Jacobian $E_L^*/(Qp_H)$. Together with $E_H^*\,dE_H^*=p_H\,dp_H$ this gives
\begin{equation}
\frac{d^3p_H}{2E_H^*}\,\frac{1}{2E_L^*}\,\frac{E_L^*}{Qp_H}=\frac{dE_H\,d\phi}{4Q} .
\label{eq:dpH}
\end{equation}
For the electron $d^3p_e/(2E_e)=\tfrac12p_e\,dE_e\,d\cos\theta_e\,d\phi_e$, and with $\mathbf k$ along $z$ the integral over $\phi_e$ gives $2\pi$. Collecting $(2\pi)^4/(2\pi)^9$ yields Eq.~\eqref{eq:lamID4}. The four variables $E_e$, $\cos\theta_e$, $E_H$ and $\phi$ are mapped to the unit hypercube with the same energy limits as above and sampled with scrambled Sobol sequences; points with $|\cos\alpha|>1$ or $E_L^*<m_L^*$ do not contribute. For the width test the final mass is drawn from the Breit--Wigner distribution of Appendix~\ref{app:width}, which adds a fifth variable.

\section{Spin-summed functions $K$}
\label{app:K}
In the $C_5^A$-only limit of Eq.~\eqref{eq:GND}, the vertex $C_5^A g^{\sigma\mu}$ of Ref.~\cite{Kumar2026P1}, the function in Eq.~\eqref{eq:MK} is
\begin{equation}
K_{\Delta N}=\frac{C_5^{A\,2}}{3}\big(P_N\!\cdot\!P_\Delta+m_N^*M_\Delta^*\big)\left[\ell\!\cdot\!\kappa+\frac{2\,(P_\Delta\!\cdot\!\ell)(P_\Delta\!\cdot\!\kappa)}{M_\Delta^{*2}}\right] ,
\label{eq:KDN}
\end{equation}
where $P_\Delta$ and $P_N$ are the kinetic momenta of the $\Delta$ and nucleon. The second term in the brackets comes from the term $-2P_\Delta^\sigma P_\Delta^\tau/(3M_\Delta^{*2})$ of the spin-$\tfrac32$ projector~\cite{HNV2007}. It is part of the on-shell spin sum: the vector-spinors built from Clebsch--Gordan combinations of polarization vectors and Dirac spinors reproduce the full projector, including this term, to $10^{-13}$. It is therefore not a spin-$\tfrac12$ admixture and cannot be dropped. Equation~\eqref{eq:KDN} is symmetric under the exchange of $\ell$ and $\kappa$; the full vertex of Eq.~\eqref{eq:GND} is not, because of the interference of its vector and axial parts, and its $K$ is evaluated for each crossing with the assignments of Eq.~\eqref{eq:cross}, which hold because $K$ is linear in each lepton momentum.

For the $\Delta\to\Delta'$ transitions we define $a=P_1\!\cdot\!P_2$, $b=P_1\!\cdot\!\ell$, $c=P_1\!\cdot\!\kappa$, $d=P_2\!\cdot\!\ell$, $e=P_2\!\cdot\!\kappa$ and $f=\ell\!\cdot\!\kappa$, and write $M_1$, $M_2$ for the effective masses and $M_{12}=M_1^2M_2^2$. The three parts of Eq.~\eqref{eq:KDD} are
\begin{align}
\frac{t_0}{64}={}&-\frac49\,af+\frac49\,\frac{a^2(be+cd)}{M_{12}}-\frac49\,\frac{a^2f}{M_1M_2}+\frac49\,bc\,\frac{M_2}{M_1}\nonumber\\
&+\frac{10}{9}\,(be+cd)+\frac49\,de\,\frac{M_1}{M_2}-\frac{10}{9}\,fM_1M_2 ,
\label{eq:t0}\\
\frac{t_1}{64}={}&-\frac89\,\frac{a^2(be-cd)}{M_{12}}-\frac43\,(be-cd) ,
\label{eq:t1}\\
\frac{t_2}{64}={}&-\frac49\,af+\frac49\,\frac{a^2(be+cd)}{M_{12}}+\frac49\,\frac{a^2f}{M_1M_2}-\frac49\,bc\,\frac{M_2}{M_1}\nonumber\\
&+\frac{10}{9}\,(be+cd)-\frac49\,de\,\frac{M_1}{M_2}+\frac{10}{9}\,fM_1M_2 .
\label{eq:t2}
\end{align}
The vector part $t_0$ and the axial part $t_2$ differ in the sign of the terms that change sign under $M_2\to-M_2$, as for the Dirac function of Eq.~\eqref{eq:KVA}, and the interference term $t_1$ is antisymmetric under $\ell\leftrightarrow\kappa$. In the static limit, $\mathbf p_1=\mathbf p_2=0$ and $M_1=M_2$, these expressions give the strengths $c_V^2$ and $\tfrac53c_A^2$ of Eq.~\eqref{eq:sigD}. We checked Eqs.~\eqref{eq:KVA}, \eqref{eq:KDN} and~\eqref{eq:t0}--\eqref{eq:t2} in each crossed form against explicit numerical sums over Dirac spinors and Rarita--Schwinger vector-spinors, built from Clebsch--Gordan combinations of polarization vectors and Dirac spinors, at random relativistic kinematics; the agreement is at the level of $10^{-10}$.

\section{Width and broadening}
\label{app:width}
\textit{Nucleon width approximation.} In the NWA~\cite{AHZ2024nwa} the opacity is averaged over the effective masses of the initial and final baryons, each distributed according to a Breit--Wigner function of full width $W$ centered at the quasiparticle value $m^*$. We sample each mass as $m=m^*+\tfrac12W\sinh t$, for which the Breit--Wigner measure becomes $dt/(\pi\cosh t)$, and restrict $|m-m^*|\leq0.95\,m^*$. To sample both the core and the tails efficiently, $t$ is drawn from an equal mixture of the Breit--Wigner distribution and a uniform distribution, and each sample is weighted by the ratio of the Breit--Wigner density to the mixture density. With $W\to0$ the quasiparticle result is recovered.

\textit{Neutral-current energy broadening.} Equation~\eqref{eq:SG} is evaluated by drawing $\varepsilon$ from the Gaussian $g_\sigma$, with $\sigma$ the standard deviation corresponding to the full width $T^2/T_W$, and shifting the energy of the scattered neutrino by $\varepsilon$ while the baryon occupations stay on shell. The factor $e^{-\varepsilon/2T}$ is carried as a weight, and the normalization is
\begin{equation}
\int d\varepsilon\,g_\sigma(\varepsilon)\,e^{-\varepsilon/2T}=e^{\sigma^2/8T^2} .
\label{eq:norm}
\end{equation}
In a dilute, nondegenerate gas, where the unbroadened response is a narrow peak, the broadened opacity increases by the analytic factor $(1+\sigma^2/2TE)^2$ to leading order; the calculation reproduces this to 2--3\%.

\textit{Rejected alternative.} We first tried Lorentzian spectral functions for the baryons in the occupation numbers. With the mean-field chemical potentials held fixed this changes the baryon densities by amounts that are not small at the widths considered, and a consistent treatment would require a self-consistent recalculation of the equilibrium. The energy broadening of Eq.~\eqref{eq:SG} avoids this.

\section{Benchmark against Ref.~\cite{Chen2007}}
\label{app:chen}

\begin{figure*}[tbp]
\includegraphics[width=\textwidth]{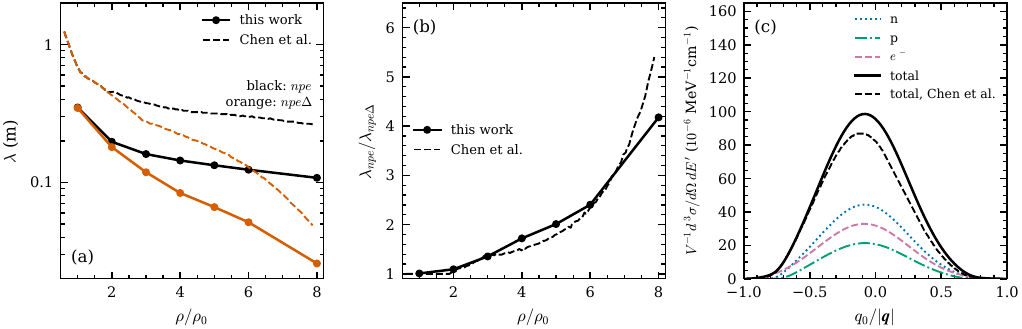}
\caption{Benchmark against Ref.~\cite{Chen2007} (GM3, $g_{\sigma\Delta}=1.3\,g_{\sigma N}$, $T=30$~MeV, $Y_L=0.4$, $E=\mu_{\nu_e}$). (a) Neutral-current mean free path in $npe$ (black) and $npe\Delta$ (orange) matter, including scattering on electrons: this work (solid with dots) and Ref.~\cite{Chen2007}, digitized from their Fig.~6 (dashed). (b) Ratio of the two mean free paths. (c) Differential cross section per volume at $4\rho_0$ in $npe$ matter, $E=226$~MeV, $|\mathbf q|=113$~MeV (as labeled in their Fig.~5), for $n$, $p$, $e^-$ and their sum (this work), and the total of Ref.~\cite{Chen2007}, digitized from their Fig.~5 (dashed). The responses and the $\Delta$ effect agree; the absolute mean free paths differ by a factor of \chenFactor.}
\label{fig:chen}
\end{figure*}

Earlier calculations of opacities of $\Delta$-admixed matter were reported in Refs.~\cite{Chen2007,ChenGuo2008,Chen2009}; those of Refs.~\cite{Chen2007,Chen2009} used the GM3 model with $g_{\sigma\Delta}=1.3\,g_{\sigma N}$, $g_{\omega\Delta}=g_{\omega N}$ and $g_{\rho\Delta}=g_{\rho N}$, $T=30$~MeV (Ref.~\cite{Chen2007} also $T=0$) and, for trapped neutrinos, $Y_L=0.4$. As an external benchmark we reproduced their setting: we implemented GM3 with the parameters listed in Ref.~\cite{Chen2007} (it gives $m^*/M=0.780$ and $E/A=-16.3$~MeV at $n=0.153$~fm$^{-3}$, as quoted there), their $\Delta$ couplings and neutral-current charges, and computed the neutral-current mean free path $(\sigma/V)^{-1}$ of neutrinos with $E=\mu_{\nu_e}$, including final-state blocking and scattering on electrons, as in their Fig.~6 [Fig.~\ref{fig:chen}(a)]. The $\Delta$s shorten it by factors of \chenRatios at 2, 4, 6 and $8\rho_0$ ($\rho_0=0.153$~fm$^{-3}$); from their figure, digitized, we obtained \chenTheirRatios. The $\Delta$ effect therefore agrees [Fig.~\ref{fig:chen}(b)]. Our absolute mean free paths in nucleonic matter, \chenLam~m, are shorter than theirs by a factor of \chenFactor (1.7 in neutrino-free matter at $4\rho_0$ for $E=100$~MeV). To locate the difference we computed the differential cross section $V^{-1}d^3\sigma/d\Omega\,dE'$ at the kinematics of their Fig.~5 ($4\rho_0$, $npe$; neutrino-free matter with $E=100$~MeV and $|\mathbf q|=50$~MeV, trapped matter with $E=\mu_{\nu_e}$ and $|\mathbf q|=113$~MeV) by a direct phase-space integration, written independently of our opacity code and checked against the heavy-target limit to 0.05\%. Our peak heights exceed theirs by 7--14\% for $n$, $p$ and the total (25\% for $e^-$ in neutrino-free matter), and the areas under the total curves by 8 and 14\%, far less than the factor in the mean free path [Fig.~\ref{fig:chen}(c)]. Integrating the same differential cross section over $q_0$ and $|\mathbf q|$ gives 0.197 and 0.145~m at 2 and $4\rho_0$, the values of our opacity code. The responses therefore agree, and the factor enters between their Fig.~5 and their Fig.~6, in the integration of their Eq.~(19), whose numerical implementation is not described in Ref.~\cite{Chen2007}.

\section{Numerical convergence}
\label{app:conv}
The energy integral in Eq.~\eqref{eq:lamR} is evaluated with 48 Gauss--Legendre points on $[\max(0,\mu-30T),\,\mu+25T]$, where $\mu$ is the chemical potential of the species. Replacing it by 24 points on $[0,\mu+25T]$ changes $\lambda_R$ by less than 3\%.

The quasi-Monte Carlo error of the four-dimensional integral is largest for neutrino absorption at low temperature, where the integrand is concentrated near the Fermi surfaces. For nucleonic matter at $3n_0$, $Y_L=0.3$ and $T=7.5$~MeV, independent scramblings give $\lambda_R$ with a scatter of $\pm3\%$ at $2^{13}$ points, $\pm1.5\%$ at $2^{15}$ points and $\pm1\%$ at $2^{17}$ points, and the mean drifts upward with the number of points until it converges at $2^{17}$--$2^{18}$ points, 5\% above the $2^{13}$-point value. In $\Delta$-admixed matter the error is smaller. The curves of Fig.~\ref{fig:ratio} use $2^{16}$ points, for which the error of $\lambda_R$ is below 3\%. The grid of Fig.~\ref{fig:scaling} and Table~\ref{tab:fits} uses $2^{13}$ points, so at $T\leq10$~MeV the neutrino ratios there may be too low by up to 5\%, which is smaller than the rms deviations of the fits. For antineutrinos, and for neutrinos at $T\geq12.5$~MeV, the errors are below 1\%. The width calculations and the sensitivity study use $2^{14}$ points. The three-body opacity uses $2^{14}$ points throughout; at $2n_0$, $Y_L=0.3$, $T=10$~MeV and $E=15$~MeV it changes by 0.5\% between $2^{14}$ and $2^{18}$ points.

\bibliography{refs}

\end{document}